%% file: main.tex
\documentclass[reprint,
superscriptaddress,
pra,
nobibnotes,
longbibliography
]{revtex4-1}

\usepackage[hyperindex,breaklinks,colorlinks]{hyperref}
\usepackage{graphicx}
\usepackage{dcolumn}
\usepackage{bm}
\usepackage{comment}
\usepackage{upgreek}
\usepackage[american]{babel}
\usepackage[usenames,dvipsnames]{color}
\usepackage[mathlines]{lineno}
\usepackage{multirow}
\usepackage{amssymb,amsmath,amsthm,enumitem}

\definecolor{bluegray}{RGB}{40,180,160}
\definecolor{navygray}{RGB}{110,140,170}
\definecolor{meadowgreen}{RGB}{0,128,0}
\definecolor{magenta}{RGB}{255,0,255}
\hypersetup{
citecolor={bluegray}, 
linkcolor={navygray}, 
}
\usepackage[utf8]{inputenc}
\DeclareUnicodeCharacter{2212}{-}
\PassOptionsToPackage{linktocpage}{hyperref}

\begin{document}

\title{Reducing~the~impact~of~radioactivity~on~quantum~circuits~in~a~deep-underground~facility}

\author{L.~Cardani}
\email{laura.cardani@roma1.infn.it}
\affiliation{INFN~Sezione~di~Roma,~00185~Roma,~Italy}

\author{F.~Valenti}
\thanks{First two authors contributed equally.}
\affiliation{PHI,~Karlsruhe~Institute~of~Technology,~76131~Karlsruhe,~Germany}
\affiliation{IPE,~Karlsruhe~Institute~of~Technology,~76344~Eggenstein-Leopoldshafen,~Germany} 

\author{N.~Casali}
\affiliation{INFN~Sezione~di~Roma,~00185~Roma,~Italy}

\author{G.~Catelani}
\affiliation{JARA~Institute~for~Quantum~Information,~Forschungszentrum~J\"ulich,~52425~J\"ulich,~Germany}

\author{T.~Charpentier}
\affiliation{PHI,~Karlsruhe~Institute~of~Technology,~76131~Karlsruhe,~Germany}

\author{M.~Clemenza}
\affiliation{Dipartimento~di~Fisica,~Universit\`{a}~di~Milano - Bicocca,~20126~Milano,~Italy}
\affiliation{INFN~Sezione~di~Milano - Bicocca,~20126~Milano,~Italy}

\author{I.~Colantoni}
\affiliation{INFN~Sezione~di~Roma,~00185~Roma,~Italy}
\affiliation{Istituto~di~Nanotecnologia,~Consiglio~Nazionale~delle~Ricerche,~c/o~Dip. Fisica,~Sapienza~Universit\`{a}~di~Roma,~00185,~Roma,~Italy}

\author{A.~Cruciani}
\affiliation{INFN~Sezione~di~Roma,~00185~Roma,~Italy}

\author{L.~Gironi}
\affiliation{Dipartimento~di~Fisica,~Universit\`{a}~di~Milano - Bicocca,~20126~Milano,~Italy}
\affiliation{INFN~Sezione~di~Milano - Bicocca,~20126~Milano,~Italy}

\author{L.~Gr{\"u}nhaupt}
\affiliation{PHI,~Karlsruhe~Institute~of~Technology,~76131~Karlsruhe,~Germany}

\author{D.~Gusenkova}
\affiliation{PHI,~Karlsruhe~Institute~of~Technology,~76131~Karlsruhe,~Germany}

\author{F.~Henriques}
\affiliation{PHI,~Karlsruhe~Institute~of~Technology,~76131~Karlsruhe,~Germany}

\author{M.~Lagoin}
\affiliation{PHI,~Karlsruhe~Institute~of~Technology,~76131~Karlsruhe,~Germany}

\author{M.~Martinez}
\affiliation{Fundaci{\'o}n~ARAID~and~Centro~de~Astropart{\'i}culas~y~F{\'i}sica~de~Altas~Energ{\'i}as, Universidad de Zaragoza, Zaragoza 50009, Spain}

\author{G.~Pettinari}
\affiliation{Institute~for~Photonics~and~Nanotechnologies,~National~Research~Council,~00156~Rome,~Italy}

\author{C.~Rusconi}
\affiliation{INFN~Laboratori~Nazionali~del~Gran~Sasso,~67100~Assergi,~Italy}
\affiliation{Department~of~Physics~and~Astronomy,~University~of~South~Carolina,~29208~Columbia,~USA}

\author{O.~Sander}
\affiliation{IPE,~Karlsruhe~Institute~of~Technology,~76344~Eggenstein-Leopoldshafen,~Germany}  

\author{A.~V.~Ustinov}
\affiliation{PHI,~Karlsruhe~Institute~of~Technology,~76131~Karlsruhe,~Germany}
\affiliation{National~University~of~Science~and~Technology~MISIS,~119049~Moscow,~Russia}
\affiliation{Russian~Quantum~Center,~Skolkovo,~143025~Moscow,~Russia}

\author{M.~Weber}
\affiliation{IPE,~Karlsruhe~Institute~of~Technology,~76344~Eggenstein-Leopoldshafen,~Germany}  

\author{W.~Wernsdorfer}
\affiliation{PHI,~Karlsruhe~Institute~of~Technology,~76131~Karlsruhe,~Germany}
\affiliation{IQMT,~Karlsruhe~Institute~of~Technology,~76344~Eggenstein-Leopoldshafen,~Germany} 
\affiliation{Institut~N\'{e}el,~CNRS~and~Universit\'{e}~Joseph~Fourier,~Grenoble,~France}

\author{M.~Vignati}
\affiliation{INFN~Sezione~di~Roma,~00185~Roma,~Italy}
\affiliation{Dipartimento~di~Fisica,~Sapienza~Universit\`{a}~di~Roma,~00185,~Roma,~Italy}

\author{S.~Pirro}
\affiliation{INFN~Laboratori~Nazionali~del~Gran~Sasso,~67100~Assergi,~Italy}

\author{I.~M.~Pop}
\email{ioan.pop@kit.edu}
\affiliation{PHI,~Karlsruhe~Institute~of~Technology,~76131~Karlsruhe,~Germany}
\affiliation{IQMT,~Karlsruhe~Institute~of~Technology,~76344~Eggenstein-Leopoldshafen,~Germany}

\date{\today}

\begin{abstract}
As quantum coherence times of superconducting circuits have increased from nanoseconds to hundreds of microseconds, they are currently one of the leading platforms for quantum information processing. However, coherence needs to further improve by orders of magnitude to reduce the prohibitive hardware overhead of current error correction schemes. Reaching this goal hinges on reducing the density of broken Cooper pairs, so-called quasiparticles. Here, we show that environmental radioactivity is a significant source of nonequilibrium quasiparticles. Moreover, ionizing radiation introduces time-correlated quasiparticle bursts in resonators on the same chip, further complicating quantum error correction. Operating in a deep-underground lead-shielded cryostat decreases the quasiparticle burst rate by a factor fifty and reduces dissipation up to a factor four, showcasing the importance of radiation abatement in future solid-state quantum hardware.
\vspace{7.5mm}
\end{abstract}


\maketitle
\newpage 


Quantum technologies based on solid-state devices are attracting a growing interest in both academic and industrial research communities, because they offer the tantalizing prospect of engineering quantum mechanical effects by using superconducting and semiconducting building blocks reminiscent of classical integrated circuits~\cite{GU20171review, MITreview2019, BurkardReview2020}. Although a daunting technological challenge, macroscopic components such as capacitors, inductors, and Josephson junctions can be inter-connected and assembled in complex quantum circuits, as recently proven by the operation of processors consisting of tens of quantum bits (qubits)~\cite{Rigetti19qubitsOtterbach2017, Song10Qubit2017, IBM16qubitsWang2018, arute2019quantum}. While these pioneering implementations showcase the advantages of solid-state platforms, one of their main challenges for future development, increasing quantum coherence, stems from the difficulty in decoupling from various noisy environments~\cite{MITreview2019}; be that dielectric defects, magnetic moments, trapped charges and vortices, spurious electromagnetic modes, or excess quasiparticles (QPs).

Quasiparticles, which can be viewed as broken Cooper pairs, degrade the performance of superconducting circuits in two ways~\cite{CatelaniQP2011}: their presence introduces dissipation, and fluctuations in their numbers give rise to noise. Although QPs are particularly damaging in circuits employing the high kinetic inductance of Cooper pairs~\cite{devisser2011number, Grunhaupt2018QPgral, grunhaupt2019granular}, often constituting the dominant source of decoherence, we will argue below that QPs can be an indicator of a more generally damaging noise source for solid-state hardware, namely radioactivity. Even a single interaction with a gamma ray, or a muon, can release up to $\sim1$~MeV in the device substrate~\cite[Section~33]{Tanabashi:2018oca}, in the form of high-energy phonons. This energy is orders of magnitude larger than the semiconducting or superconducting gap of device materials. While the resulting QPs constitute the signal for several types of superconducting detectors used in particle physics and astrophysics~\cite{day2003broadband, Swenson2010, Moore2012, Zmuidzinas2012Review, Ulbricht2015Xray}, radioactivity has so far received little attention as a possible source of decoherence in solid state quantum hardware.

\begin{figure*}[!t]
\begin{center}
\def\svgwidth{1 \textwidth} 
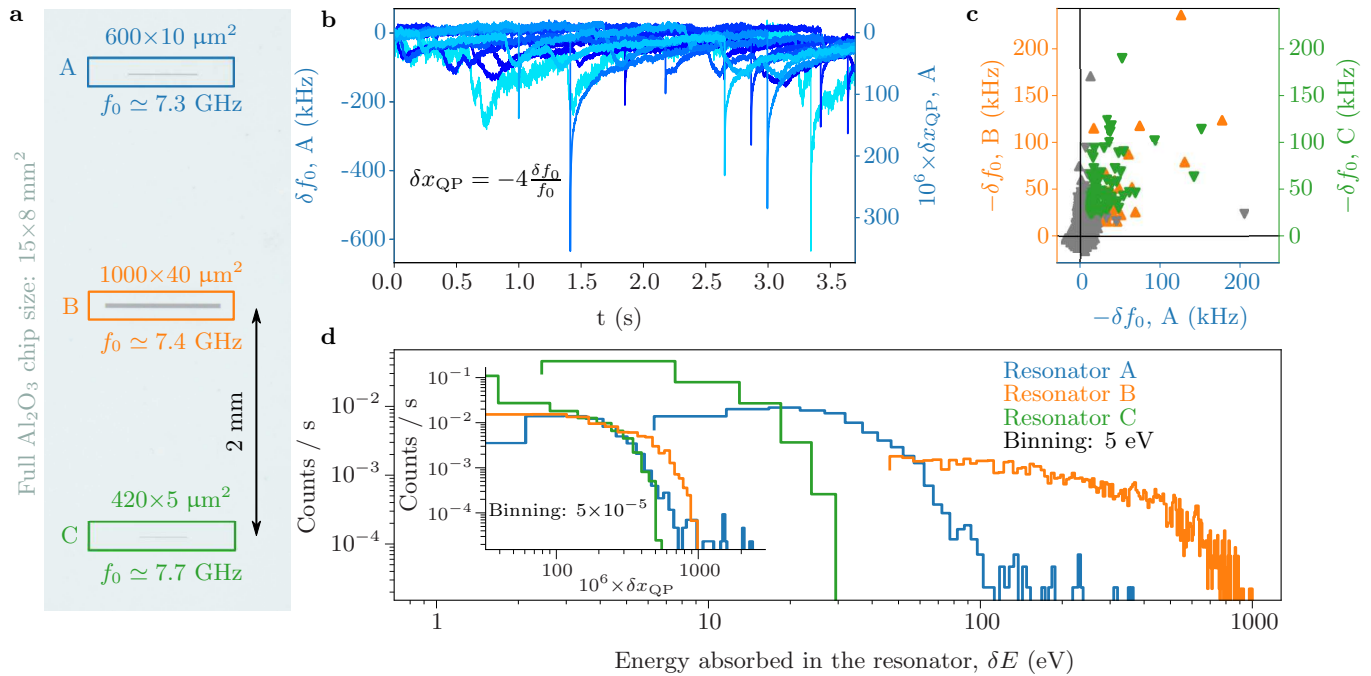
\caption{ \textbf{Quasiparticle bursts and deposited energy in grAl resonators.} \textbf{a}, 
Photograph of the central part of the sapphire chip, supporting three $20$~nm thick grAl resonators, labeled A, B, and C. \textbf{b}, Overlay of ten measured time traces for the resonant frequency shift $\delta f_0$ of resonator A. Similarly to Refs.~\cite{Swenson2010, Grunhaupt2018QPgral, wang2014QP}, quasiparticle (QP) bursts appear as sudden drops, given by the sharp rise in kinetic inductance, followed by a relaxation tail. The $y$-axis on the right hand side shows the corresponding fractional quasiparticle density shift $\delta x_\mathrm{QP} = - 4 \delta f_0 / f_0$. For clarity, the shown traces are selected to contain a QP burst; on average, only one trace in 10 contains a QP burst. To highlight the fact that QP  bursts  are  correlated in  time, in panel \textbf{c} we plot the measured frequency shifts of resonator B (upward triangles) and C (downward triangles) versus the frequency shift of resonator A. Colored markers correspond to values above threshold, with the threshold defined as two standard deviations of the baseline fluctuations (cf. Suppl. Mat.). Therefore, each colored marker depicts a time correlated QP burst between resonators A-B (orange) and A-C (green). \textbf{d}, Estimated distribution of the energy absorbed in the resonators $\delta E=\delta x_\mathrm{QP} \Delta_{\mathrm{grAl}} n_{CP} V$, calculated from the measured $\delta x_\mathrm{QP}$ shown in the inset, where $\Delta_{\mathrm{grAl}} \simeq 300$~$\upmu$eV is the grAl superconducting gap, and $n_{CP} = 4\times 10^6$~$\upmu$m$^{-3}$ is the volume density of Cooper pairs, and $V$ is the volume of each resonator. For each burst, the energy deposited in the substrate is estimated to be $10^3-10^4$ times greater than $\delta E$ (cf. Suppl. Mat.). The total QP burst rate $\Gamma_B$ is obtained by counting all bursts above the common threshold $\delta x_\mathrm{QP} = 5 \times 10^{-5}$.}
\label{fig:intro}
\end{center}
\vspace{-5mm}
\end{figure*}

Remarkably, Ref.~\cite{Vepslinen2020impact} has recently shown that the coherence limit imposed by ionizing radiation for transmon type qubits is in the millisecond range, only one order of magnitude above the state-of-the-art. Moreover, as dielectric losses are steadily decreased~\cite{Gambetta2016Dielectric, place2020Tantalum}, further improving the coherence of solid-state devices will soon hinge on the reduction of QPs, and more generally on ionizing radiation abatement. Here, we demonstrate that by reducing radioactivity we lower the internal dissipation in superconducting microwave resonators by factors two to four, and the QP burst rate by a factor fifty. This was achieved by a combination of material selection and cleaning, and by shielding under the 1.4~km granite layer at the Gran Sasso National Laboratory (L'Aquila, Italy), corresponding to a 3.6~km water equivalent.

\begin{figure*}[!htb]
\begin{center}
\def\svgwidth{1 \textwidth} 
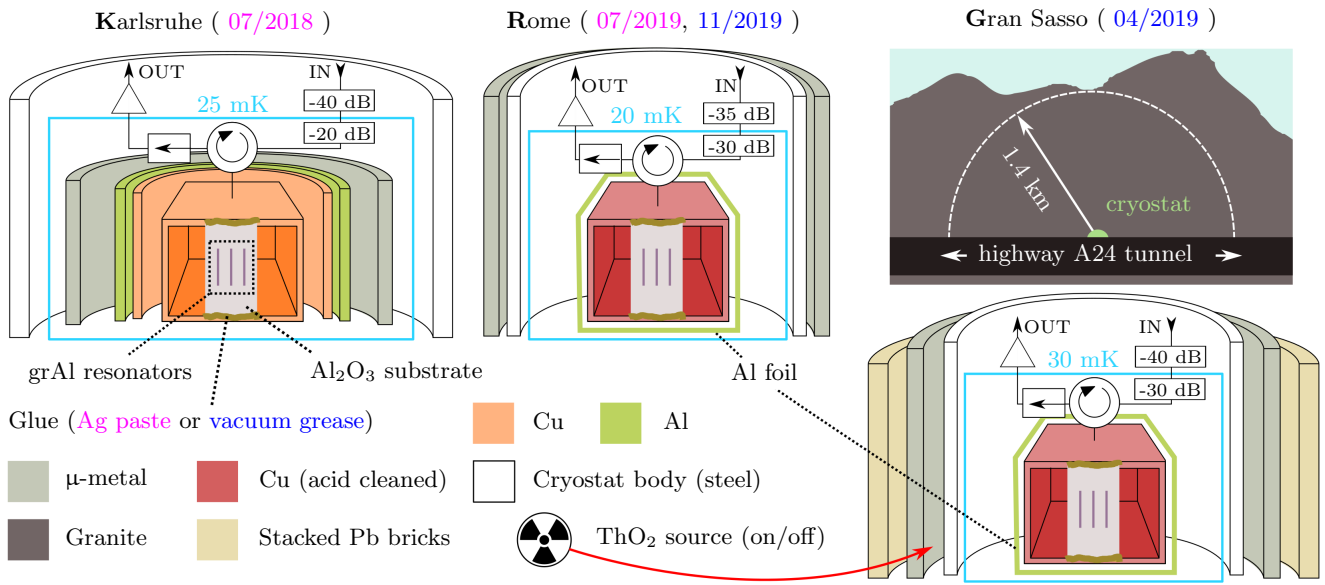
\caption{ \textbf{Three different setups with various degrees of shielding against ionizing radiation.}
Schematic half-sections of the setups, in Karlsruhe, Rome, and Gran Sasso, denoted \textbf{K}, \textbf{R}, and \textbf{G}, respectively. The measurement dates for each setup are indicated in the top labels. The sapphire chip is glued to a copper waveguide using either silver paste (K and R, magenta) or vacuum grease (G and R, blue). A circulator routes the attenuated input signal to the sample holder, and the reflected output signal to an isolator and an amplification chain (cf. Suppl. Mat.). In the R and G setups, the waveguide is etched with citric acid to remove possibly radioactive contaminants. The G setup, located under 1.4~km of granite (3.6~km water equivalent) is operated in three configurations. First, the cryostat is surrounded by a 10 cm thick wall of lead bricks. Two days later, the bricks were removed. Finally, we added a ThO$_2$ radioactive source next to the cryostat body (cf. red arrow). }
\label{fig:setup}
\end{center}
\vspace{-8mm}
\end{figure*}

In thermal equilibrium, at typical operational temperatures of 20~to~50~mK, QPs should be an extremely rare occurrence in commonly used materials such as Al and Nb, with critical temperatures well above 1~K. However, the detrimental effects of non-equilibrium QPs are routinely observed in a variety of devices~\cite{Aumentado2004, Shaw2008, barends2011minimizing, devisser2011number, Zgirski2011, Riste2013, wang2014QP, LevensonFalk2014QPnanobridge, Janvier2015, vanWoerkom2015, Gustavsson2016, Hays2018, serniak2018hot}, including the microwave resonators used in this work (cf. Fig.~\ref{fig:intro}). The multifarious QP sources include stray infrared radiation~\cite{barends2011minimizing, serniak2018hot},  high-power microwave drive~\cite{deVisserQPpower2014}, and phonons in the device substrate~\cite{Patel2017, BespalovQPmodel2017, HenriquesValenti2019} resulting from environmental or cosmic radioactivity. The latter is potentially damaging for any solid-state quantum hardware, not only superconducting, as it can give rise to correlated energy bursts in devices on the same chip. Indeed, in the case of superconducting resonators, high energy phonons in the device substrate produce correlated QP spikes orders of magnitude above the baseline~\cite{Swenson2010, Moore2012}, visible as abrupt frequency drops (see Fig.~\ref{fig:intro}b and c). Even though the rate of these bursts appears to be modest, one every few seconds ~\cite{Swenson2010, devisser2011number, Grunhaupt2018QPgral}, the ensuing relatively long-lasting and correlated effects can hinder quantum error correction protocols. \newpage

Superconducting circuits can be sensitive to a variety of radioactive sources, depending, among others, on the distance from the device, penetrating power, spectral distribution, and shielding. So-called \textit{far} sources consist of cosmic rays, mainly muons at a rate of $\sim$1~cm$^{-2}$min$^{-1}$, as well as decay products of location-specific contaminants. Even when far sources can be shielded, using e.g. Pb screens or underground facilities, \textit{near} sources such as residues from handling and machining, or radioactive isotopes in the sample holder and the sample itself might need to be mitigated by material selection and decontamination.

We use high kinetic inductance granular Aluminum (grAl) superconducting resonators (see Fig.~\ref{fig:intro}a) as a sensitive QP probe, following the principle of kinetic inductance detectors \cite{day2003broadband}. Shifts in their resonant frequency $f_0^{-1}=2\pi\sqrt{LC}$, where $C$ is the  capacitance of the mode, directly reflect changes in the inductance $\delta L/L=-(2/ \alpha) \delta f_0/f_0$, where $\alpha$ is the ratio of kinetic inductance over the total inductance. In the case of high kinetic inductance materials such as grAl, where the geometric inductance can be neglected~\cite{Grunhaupt2018QPgral}, the measured relative frequency shift informs on the corresponding change in the number of QPs with respect to the number of Cooper pairs: $\delta x_{\mathrm{QP}}=2\delta L/L=-4\delta f_0/f_0$.

The resonators were fabricated using optical lithography on a 1.2~cm$^2$ and 330~$\upmu$m thick sapphire substrate. Their dimensions and corresponding resonant frequencies $f_0$ are listed in Fig.~\ref{fig:intro}a. The measurements were performed in a magnetically shielded and infrared filtered environment \cite{Kreikebaum2016IRshielding, grunhaupt2019granular}, at low signal drive powers, in the range of $\bar{n}=1$ circulating photons, which are the typical conditions for quantum circuits. Furthermore, we used a 3D waveguide sample holder \cite{kou2018simultaneous} in order to minimize the electric field density at the interfaces and reduce coupling to dielectric losses \cite{calusine2018analysis}. Under these conditions, losses in grAl resonators are dominated by non-equilibrium QPs \cite{HenriquesValenti2019}. Using shielding and filtering against stray infrared photons \cite{barends2011minimizing,serniak2018hot}, the QP population can be reduced to levels at which the contribution from radioactivity is dominant, making grAl resonators an effective diagnostics tool. 

In Fig.~\ref{fig:intro}b we show typical time traces for the frequency shift $\delta f_0$ of resonator~A, measured in a cryostat above ground. We observe abrupt drops of $f_0$, indicative of a QP burst in the resonator film, followed by a relaxation tail, associated with QP recombination and diffusion, similarly to Refs.~\cite{Swenson2010, Moore2012, wang2014QP, Grunhaupt2018QPgral}, one every $\sim 10$~s. We interpret them as the aftermath of ionizing events in the substrate, causing an energy release in the form of phonons, which in turn produce QPs. Indeed, as shown in Fig.~\ref{fig:intro}c and in Suppl. Mat., most QP bursts in resonator A are correlated with those in resonators B and C, proving the key role played by substrate phonons \cite{Moore2012}. Notice that although resonator C is twice as far from resonator A compared to B, the correlation plot does not appear qualitatively different, indicating that in our present geometry QP bursts are time-correlated over at least 10~mm$^2$ areas of the chip, similarly to Refs.~\cite{Swenson2010,Moore2012}. The histogram of the QP burst rate as a function of the energy absorbed in the resonators is shown in Fig.~\ref{fig:intro}d. We estimate the efficiency of phonon absorption from the substrate into the resonators to be $10^{-3}-10^{-4}$, placing the energy deposited in the substrate by each ionizing impact in the keV-MeV range (cf. Suppl. Mat.).

\begin{figure}[!t]
\begin{center}
\def\svgwidth{1 \columnwidth} 
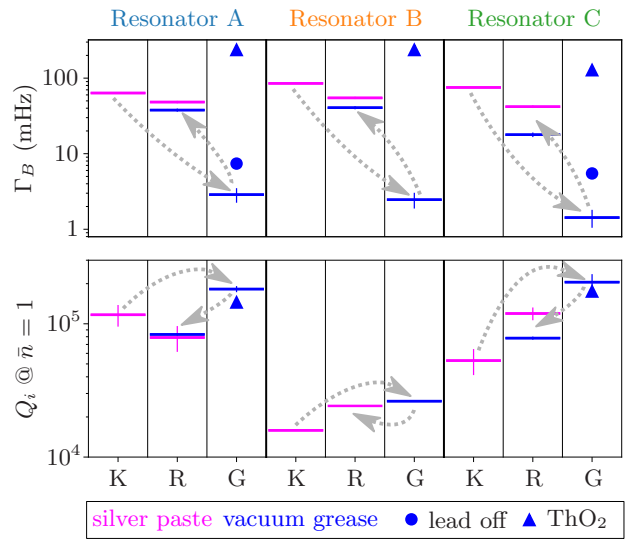
\caption{
\textbf{Effect of ionizing radiation shielding on resonator performance.} Quasiparticle burst rate ($\Gamma_B$, top) and internal quality factor at single photon drive ($Q_i$, bottom) for all resonators and setups. When the sample is cleaned and tested in the R setup, the measured $\Gamma_B$ and $Q_i$ values are comparable to those obtained in K. Nevertheless, we observe a consistent lowering of $\Gamma_B$ when using vacuum grease instead of Ag paste.
Measurements in the G setup show a reduction in QP burst rate $\Gamma _B$ (factor fifty) and dissipation (up to a factor four). In G, removing the lead shielding increases $\Gamma_B$ by a factor two. Adding a ThO$_2$ radioactive source next to the cryostat body yields a $\Gamma_B$ greater than the one measured above ground, and decreases the internal quality factor $Q_i$ by $18\pm3$\%. Error bars are not shown when smaller than the marker size. The chronological order of measurements in the three different setups is indicated by the dotted gray arrows.}
\label{fig:bursts:rates}
\end{center}
\end{figure}

In the following, we will use the QP burst rate as an indicator of the ionizing radiation flux, while we perform various combinations of material selection, cleaning, and shielding. The three setups, located in \textbf{K}arlsruhe, \textbf{R}ome, and \textbf{G}ran Sasso, denoted by K, R, and G, are schematized in Fig.~\ref{fig:setup}, and the dates of the four measurement runs are indicated by the top labels. The corresponding measured QP burst rates and internal quality factors of the resonators are listed in Fig.~\ref{fig:bursts:rates}. 

Both the K and R setups are located above ground. The K setup is typical for superconducting circuit experiments and features additional magnetic shielding compared to the R setup, consisting of a superconducting and a $\upmu$-metal barrel encasing the waveguide. In the R setup, designed to minimize the contribution to radioactivity from near sources, we cleaned the sample holder and its mounting parts with citric acid and hydrogen peroxide to reduce surface contamination, we removed the potentially radioactive indium wire used for sealing the cooper cap, and we substituted lead soldering with more radio-pure Araldite glue. To compare the results from the three setups, we performed two runs in the R setup, using either the potentially more contaminated silver paste (as in K), or radiopure vacuum grease (as in G) to glue the sapphire chip to the copper waveguide. 

In order to reduce the radioactivity contribution from far sources as well, the cleaned assembly used in the R setup is also used in the G setup. Here, the 3.6 km water equivalent of rock overburden reduces the cosmic ray flux by six orders of magnitude. In a first measurement run, the cryostat is surrounded by a wall of 10~cm~thick Pb bricks. We perform two additional measurements by first removing the bricks, and then exposing the cryostat to a $^{232}$Th source in the form of ThO$_2$.

The QP burst rate (cf. Fig.~\ref{fig:bursts:rates}, top) measured in the R setup is lower than that measured in the K setup, with a better suppression using vacuum grease instead of silver paste to glue the chip. The strongest suppression, up to a factor fifty, is achieved in the G setup, proving the preponderance of far radioactive sources. Removing the lead shielding increases the burst rate by a factor two, and adding the ThO$_2$ source increase the rate beyond above ground levels, confirming the radioactive origin of the bursts. The internal quality factors (cf. Fig.~\ref{fig:bursts:rates}, bottom) are anticorrelated with the burst rates between above and underground measurements, achieving up to a fourfold increase in the G setup. Notice that the burst rate is not simply a proxy for the quality factor, as indicated by the fact that the quality factor in the G setup only decreased by $\sim 20$\% when the QP burst rate was increased by two orders of magnitude by using the ThO$_2$ source. This shows that different radiation sources (i.e. ThO$_2$ and cosmic rays) affect the device in different ways, so more detailed studies of various sources are needed.

In conclusion, we showed that the performance of superconducting circuits at the current level of coherence can be significantly degraded by environmental radioactivity in a typical above ground setting, in particular due to ionizing interactions in the device substrate. We demonstrated that the rate of correlated quasiparticle bursts is reduced by up to a factor fifty by shielding in a deep-undeground facility and by a radioactive decontamination in the near environment of the sample. Furthermore, the quality factors of high kinetic inductance superconducting resonators improved up to a factor four with respect to above ground values.

These first observations highlight the need for a systematic assessment of radioactive sources which can produce energy bursts in solid state quantum hardware, as well as for a better understanding of the relevant chains of mechanisms, such as the creation of electron-hole pairs, and the excitation of high energy phonons, which potentially limit the performance of superconducting and semiconducting devices. The effectiveness of radiation abatement and phonon damping solutions, such as phonon traps~\cite{Karatsu2019,HenriquesValenti2019}, will determine whether the next generation of solid state quantum processors will need to be operated in deep-underground facilities.

We are grateful to M. Devoret, R. McDermott, A. Monfardini and M. Calvo for insightful discussions. We acknowledge M.~Iannone for his help in designing and fabricating the components for the microwave readout in the underground cryogenic facility, and M. Guetti for the installation of the RF cables and for the assistance in the cryogenic operations. We thank M. Perego, E. Tatananni, A. Rotilio, A. Corsi, B. Romualdi, L. Radtke, S. Diewald, and A. Lukashenko for technical support. This work was funded by INFN under Grant73-Demetra, by the European Research Council (FP7/2007-2013) under contract CALDER no.~335359, by the Alexander von Humboldt foundation in the framework of a Sofja Kovalevskaja award endowed by the German Federal Ministry of Education and Research, and by the Initiative and Networking Fund of the Helmholtz Association, within the Helmholtz future project scalable solid state quantum computing. AVU acknowledges partial support from the Ministry of Education and Science of Russian Federation in the framework of the Increase Competitiveness Program of the National University of Science and Technology MISIS (Grant No. K2-2017-081). This work makes use of the Arby software for Geant4-based Monte Carlo simulations, that has been developed in the framework of the Milano~-~Bicocca R\&D activities, maintained by O.~Cremonesi and S.~Pozzi.

\bibliography{refe_full_nourl}

\newpage
\onecolumngrid
\newpage

\section*{{Supplementary Material}}
\hrulefill

\section{Schematics of microwave setups}\label{app_schematics}
\begin{figure*}[h!]
\begin{center}
\def\svgwidth{1 \textwidth}  
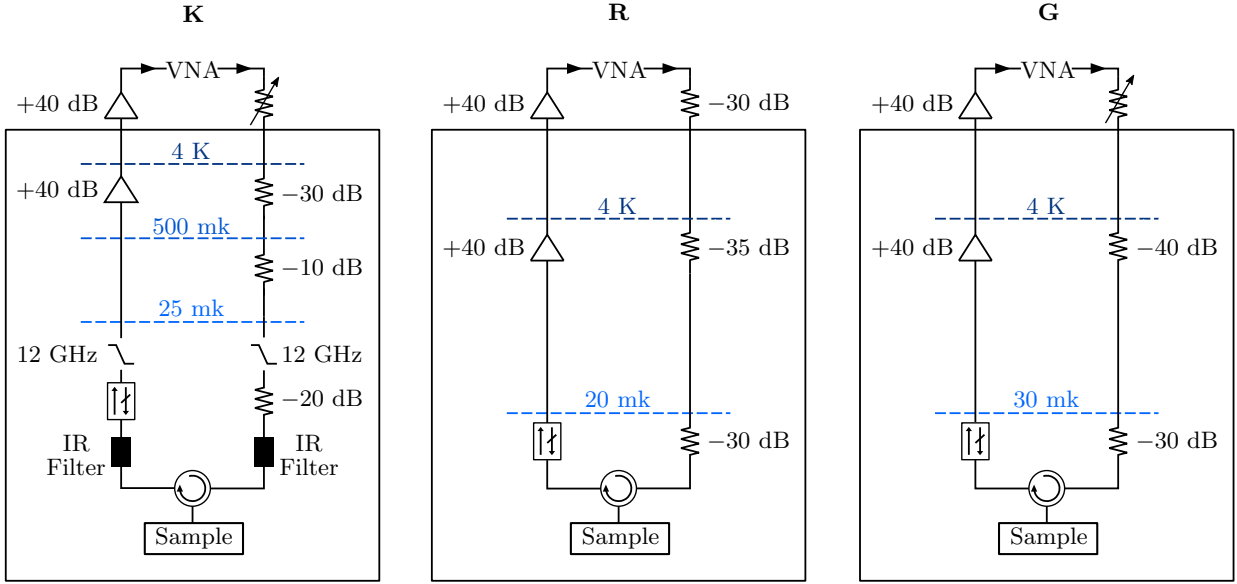
\caption{\textbf{Schematics of the input and output lines of the K, R and G setups}. The displayed components are thermalized to the nearest temperature stage indicated above them.
}\label{A_fig_schematics}
\end{center}
\end{figure*}

We show schematics of the microwave wiring for the K, R, and G setups in Fig.~\ref{A_fig_schematics}. The three setups are similar, with the notable difference that K is equipped with microwave and IR filters.
\vspace{3 mm}
\vspace{10 mm}

\newpage
\onecolumngrid
\newpage
\section{Sample mounting and cryostat shielding}
\begin{figure*}[h!]
\begin{center}
\def\svgwidth{1 \textwidth}  
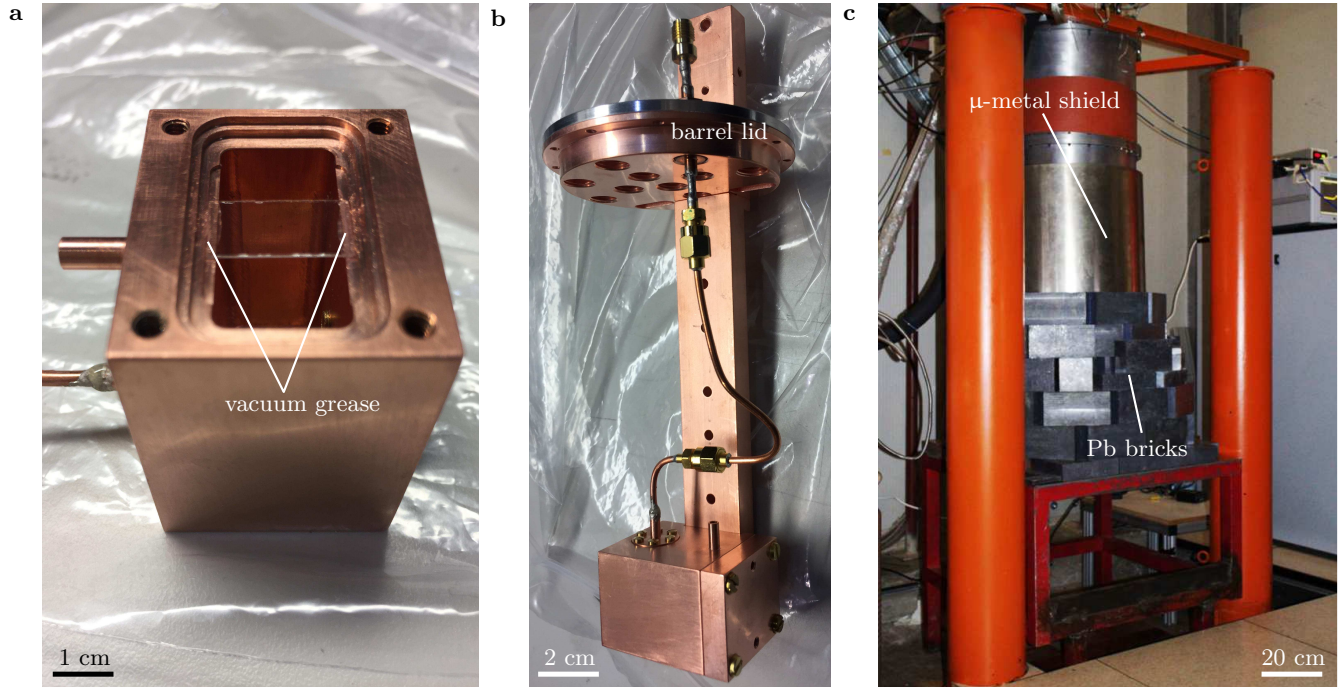
\caption{\textbf{Sample mounting and cryostat shielding}. \textbf{a}, Copper waveguide without the cap. The sapphire chip is glued to the waveguide with vacuum grease (shown) or Ag paste. \textbf{b}, Mounting copper rod with capped waveguide screwed on. In the K setup, a Cu/Al bilayer barrel and a $\upmu$-metal barrel are screwed onto the lid, encapsulating the rod-waveguide ensemble. \textbf{c}, Dilution cryostat in the G setup. Notice the $\sim$2 mm thick $\upmu$-metal barrel and the wall of $\sim 20\times10\times5$~cm$^3$ Pb bricks surrounding it.
}\label{A_fig_cryostat}
\end{center}
\end{figure*}

We show the sample mounting and the dilution cryostat of the G setup in Fig.~\ref{A_fig_cryostat}.
\vspace{3 mm}
\vspace{10 mm}

\newpage
\section{Frequency drift of the resonators over time}\label{app_drift}
\begin{figure*}[h!]
\begin{center}
\def\svgwidth{1 \textwidth}  
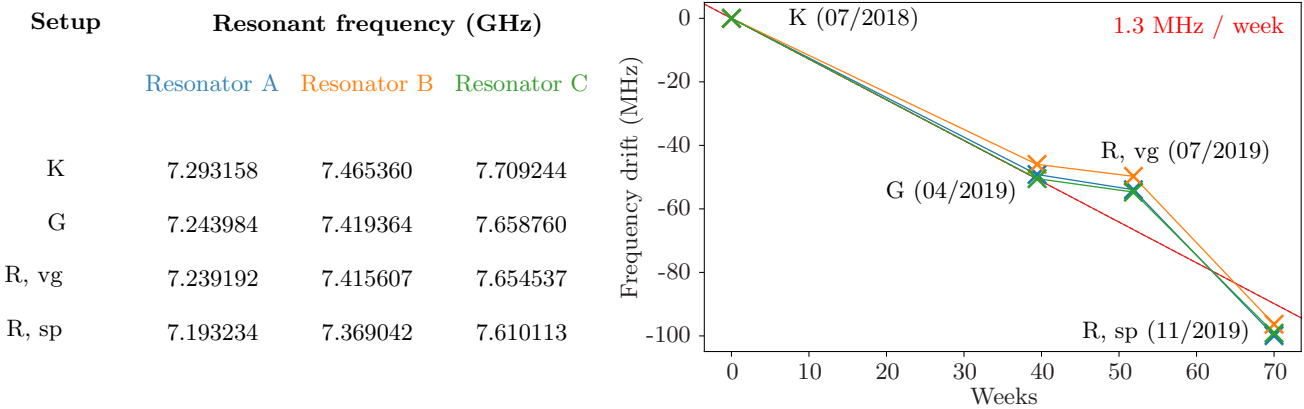
\caption{\textbf{Frequency drift}. Time evolution of the resonant frequencies of resonators A, B, and C over the four measurements runs. As a reference, we plot a linear negative drift of 1.3~MHz per week (red dashed line).
}\label{A_fig_drift}
\end{center}
\end{figure*}

We report the frequency drift over time of the three grAl resonators in Fig.~\ref{A_fig_drift}. During a cooldown, the resonator frequency does not change; between cooldowns the chip is stored at room temperature and atmospheric pressure.
\vspace{5 mm}
\vspace{1 mm}

\newpage 
\section{Measurement of correlated bursts with a VNA}\label{app_corr}

\begin{figure*}[h!]
\begin{center}
\def\svgwidth{1 \textwidth}  
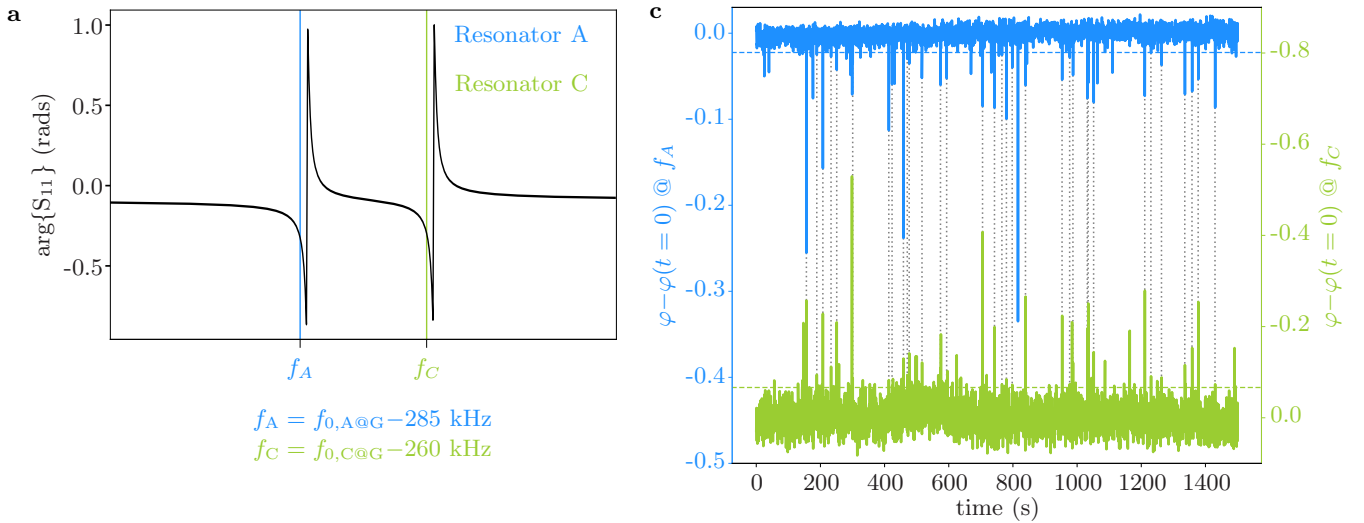
\caption{\textbf{Time multiplexed measurement of the phase response of two resonators using a Vector Network Analyzer (VNA)}. \textbf{a}, The phase response of the reflected signal vs. frequency (black line) is sampled at two points in the vicinity of the resonant frequency of the two resonators (blue and green). \textbf{b}, Time multiplexed phase response $\varphi_{A,C} (t)$ = $\mathrm{arg}\{S_{11}\}(f_{A,C},t)$ for resonators A and C vs. time in G with the ThO$_2$ source. For clarity, we invert the sign of the right-hand y axis. The time interval between measurements at the two different frequencies is 10~ms. Bursts exceeding the two standard deviation threshold (horizontal dashed lines) for both resonators are highlighted by vertical dotted gray lines. The relaxation tail after the QP bursts is not resolved due to the fact that the time interval between successive measurements in each trace is relatively long, $\Delta t \approx 0.3$~s (see text for details).
}\label{A_fig_corr}
\end{center}
\end{figure*}

In order to investigate whether QP bursts are correlated in time, we measure time multiplexed traces of the phase response of two resonators by employing a Keysight E5071C Vector Network Analyzer (VNA). The method is summarized in Fig.~\ref{A_fig_corr}. The time needed to switch between the two frequencies is the inverse of the IF bandwidth, $1/100$~Hz~$=10$~ms. The time interval $\Delta t$ between each pair of measured points is dominated by the time needed to transfer their values from the VNA to the measurement PC. We estimate it by dividing the total acquisition time by the number of acquired 2-point measurements, giving an effective sampling period $\Delta t \approx 0.3$~s.

As discussed in the main text, the relaxation time after a QP burst is from tens to hundreds of milliseconds, depending on the setup. This time is short compared to the sampling period $\Delta t$, which implies that the QP bursts are sampled by at most two points along their relaxation tail. This method, while it has the advantage of being conceptually simple and possible to implement with off-the-shelf electronics, is restricted to measuring only two resonators at a time, and is clearly limited when it comes to evaluating the size of QP bursts and their exact position in time. These limitations can be overcome by using frequency domain multiplexing and custom designed electronics, similarly to Refs.~\cite{Swenson2010, Moore2012,  bourrion2012electronics, hickish2016review}.
\vspace{5 mm}
\vspace{10 mm}

\newpage
\section{Measurement of the internal quality factor}
\begin{figure*}[h!]
\begin{center}
\def\svgwidth{1 \textwidth}  
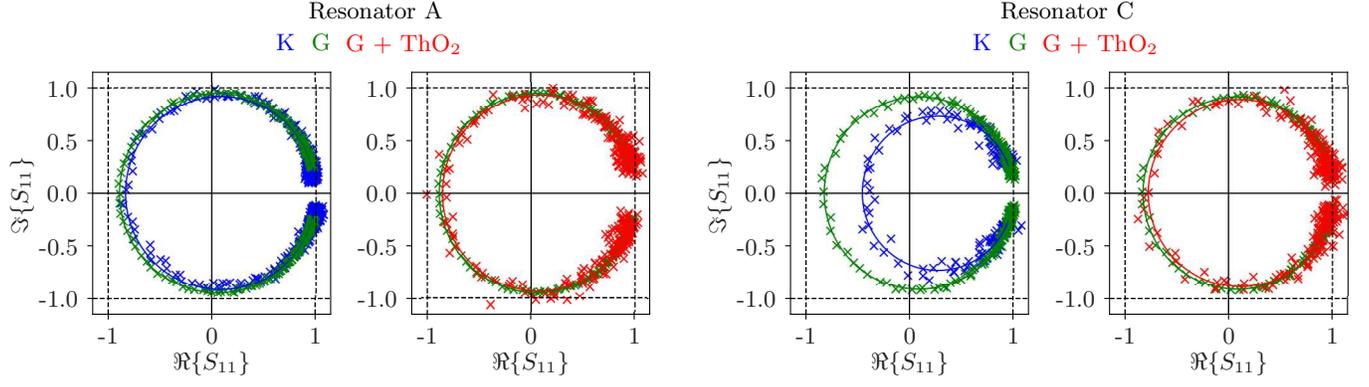
\caption{\textbf{Measurement of the internal quality factor.} Reflection coefficient of resonator A (left) and C (right), normalized to the sample holder response and plotted in the complex plane. We show data for both resonators in G (green) compared to those in K (blue) and in G with the Pb shield removed and ThO$_2$ source added (red). Crosses and solid lines indicate the raw data and the circle fit from Ref.~\cite{shahid2011reflection}, respectively.
}\label{A_fig_qi}
\end{center}
\end{figure*}

We fit the complex reflection coefficient $S_{11}$ of the resonators with the procedure detailed in Ref.~\cite{shahid2011reflection} in order to extract the internal and coupling quality factors and the resonant frequency. We compute the average number of drive photons circulating in the resonators as $\Bar{n} = 4 P_\mathrm{cold} Q_l^2 / (\hbar \omega_0^2 Q_c )$, where $P_\mathrm{cold}$ is the VNA probe power minus the nominal attenuation on the line down, $Q_l$ and $Q_c$ are the loaded and coupling quality factors, and $\omega_0$ is the resonant frequency in radians per second. We set the IF bandwidth of the VNA to 10~kHz and we average for 500 times. We plot the results in Fig.~\ref{A_fig_qi}.
\vspace{5 mm}
\vspace{10 mm}

\newpage 
\section{Choosing the bin size for the QP bursts energy distribution}\label{none}
\begin{figure*}[h!]
\begin{center}
\def\svgwidth{1 \textwidth}  
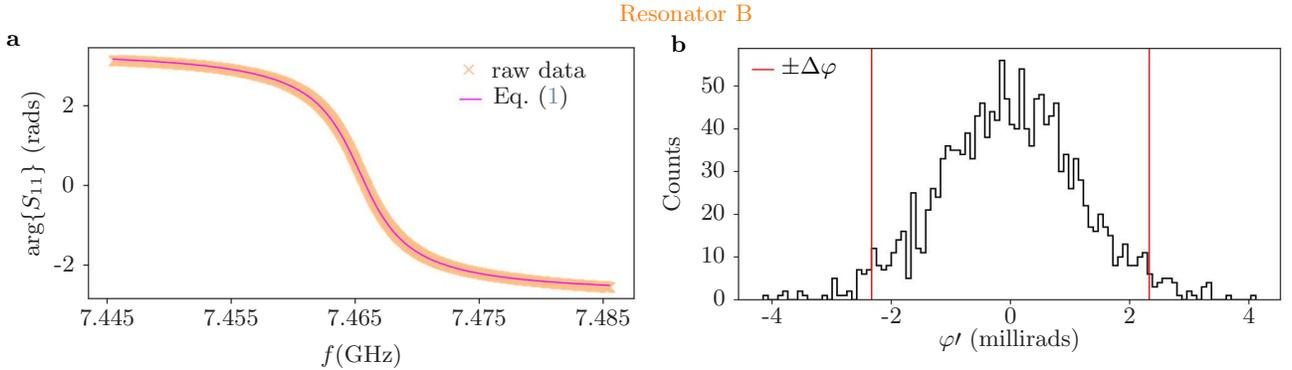
\caption{\textbf{Phase response vs. frequency and phase uncertainty for resonator B}. \textbf{a}, Measured phase response vs. frequency (orange crosses) fitted with Eq.~\eqref{arctanfit} (magenta line). \textbf{b} Distribution of the nearest point differences $\varphi'$ of the measured phase response with the arctangent component removed.
}\label{A_fig_uncert}
\end{center}
\end{figure*}

In Fig.~1d in the main text, we show a distribution of QP bursts vs. the energy absorbed by the resonators. In the following we describe the method used to define a common bin size $\Delta E$ for the distribution of QP bursts. The goal is to find the smallest $\Delta E$ that is larger than the uncertainty corresponding to the phase response uncertainty $\Delta \varphi$ for all resonators.
The phase response of the reflected signal vs. frequency follows the form
\begin{equation}\label{arctanfit}
    \varphi ( f ) = \arctan \left( 2 Q_l \frac{f_0-f}{f_0} \right),
\end{equation} 
where $f_0$ is the resonant frequency, and $Q_l$ is the loaded quality factor (cf.~Fig.~\ref{A_fig_uncert}a). The response around $f_0 + \delta f_0$ is linear for $\delta f_0 \to 0$ and asymptotic to $\pm \pi$ for $\delta f_0 \to \mp \infty$: the greater the shift $\delta f_0$ corresponding to large QP bursts, the greater the frequency uncertainty corresponding to $\Delta \varphi$. Thus, we can only quantitatively trust shifts $\delta f_0$ up to the point where the phase uncertainty $\Delta \varphi$ corresponds to a frequency difference smaller than the bin size in frequency units, $\Delta f_0$. This places the following constraints:
\begin{align}
& \left | \varphi ( f_0 + \delta f_0) - \varphi ( f_0 +  \Delta f_0 + \delta f_0) \right | < \Delta  \varphi \nonumber  \\ 
  \iff   & \left | \varphi \left( f_0 +  \delta E \frac{f_0}{4 V \Delta_0 n_\mathrm{CP}} \right) - \varphi \left( f_0 +  \Delta E \frac{f_0}{4 V \Delta_0 n_\mathrm{CP}} +  \delta E \frac{f_0}{4 V \Delta_0 n_\mathrm{CP}} \right) \right | < \Delta \varphi, \label{eq_condition}
\end{align}
where $\Delta f_0$ and $\Delta E$ are the bin sizes in frequency and energy units respectively, $\delta E = \delta x_\mathrm{QP} n_\mathrm{CP} \Delta_0 V$ is the burst in energy units and $\delta x_\mathrm{QP} = -4 \delta f_0 / f_0$ is the burst in fractional QP density units, $V$ is the volume of the resonator, $\Delta_0 \simeq 300\;\upmu$eV is the superconducting gap of grAl, $n_\mathrm{CP} = 4 \times 10^6\;\upmu$m$^{-3}$ is the Cooper pair density of Al, and $f_0$ is the unperturbed resonant frequency. 

In order to calculate the uncertainty of the response $\Delta \varphi$, we substract the fitted arctangent dependence of Eq.~\eqref{arctanfit} from the measured arg\{$S_{11}$\}   (cf.~Fig.~\ref{A_fig_uncert}a), and we compute its nearest point difference $\varphi'$. We define  $\Delta \varphi = 2\; \mathrm{stdev}(\varphi ' )$  (cf.~Fig.~\ref{A_fig_uncert}b), and we find $\Delta \varphi \lesssim 10$~milliradians for all resonators.

Finally, using $\Delta E=5$~eV, Eq.~\ref{eq_condition} is satisfied for $\Delta \varphi = 10$~milliradians and for all measured QP burst in all resonators.
\vspace{2 mm}

\newpage 
\section{Distribution of QP bursts in R and G setups}\label{app_dist}
We report the distribution of QP bursts for resonators in R and G, to complement the distribution for resonators in K shown in the main text (Fig.~1d, inset). We omit the distribution of bursts in G with the Pb shield present, resulting in the lowest burst rate (one every 10 minutes) due to the lack of sufficient statistics.
\vspace{10 mm}
\begin{figure*}[h!]
\begin{center}
\def\svgwidth{1 \textwidth}  
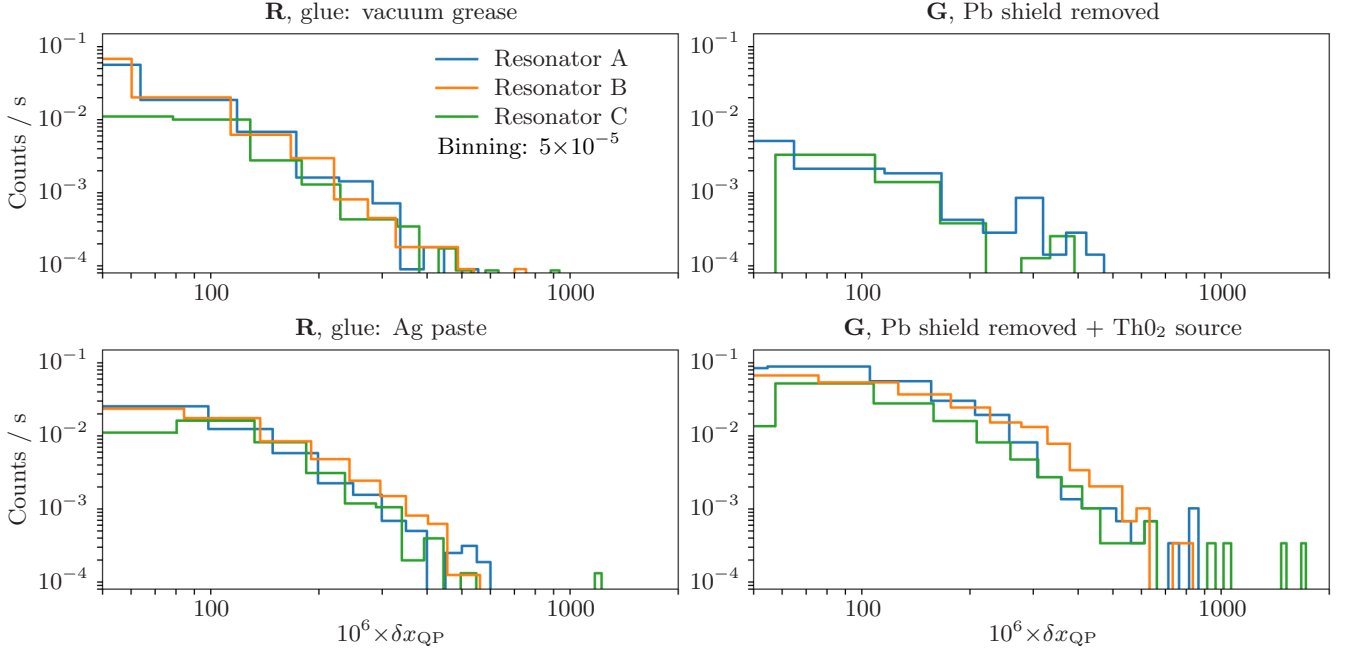
\caption{\textbf{Distribution of QP bursts in R and G}. The bin size is $5\times 10^{-5}$ in all panels. For the G setup with the Pb shield removed, data were not available for resonator B due to technical and time constraints. 
}\label{A_fig_histograms}
\end{center}
\end{figure*}
\vspace{10 mm}

\newpage 
\section{Phonon absorption efficiency}\label{efficiency}
The Cooper pair condensate of a resonator absorbs only a small fraction $\epsilon$ of the total energy deposited in the substrate chip. This fraction depends, among others, on the material and volume of the superconductor, its interface with the substrate, the material and geometry of the substrate and its coupling to the thermal bath. As a consequence, the efficiency of energy absorption cannot be calculated a priori but requires a dedicated measurement.

To our knowledge, there are no published results on the absorption efficiency of a grAl film deposited on a sapphire substrate. Due to the phonon acoustic mismatch between grAl and the sapphire substrate, we expect the efficiency $\epsilon$ to be very small.
To constrain this parameter without contaminating the prototype, we exposed the chip to a removable ThO$_2$ source and compared the calculated spectrum of the energy deposited in the substrate with the measured spectrum of the energy absorbed by the three resonators. 

\begin{figure*}[h!]
\begin{center}
\def\svgwidth{1 \textwidth}  
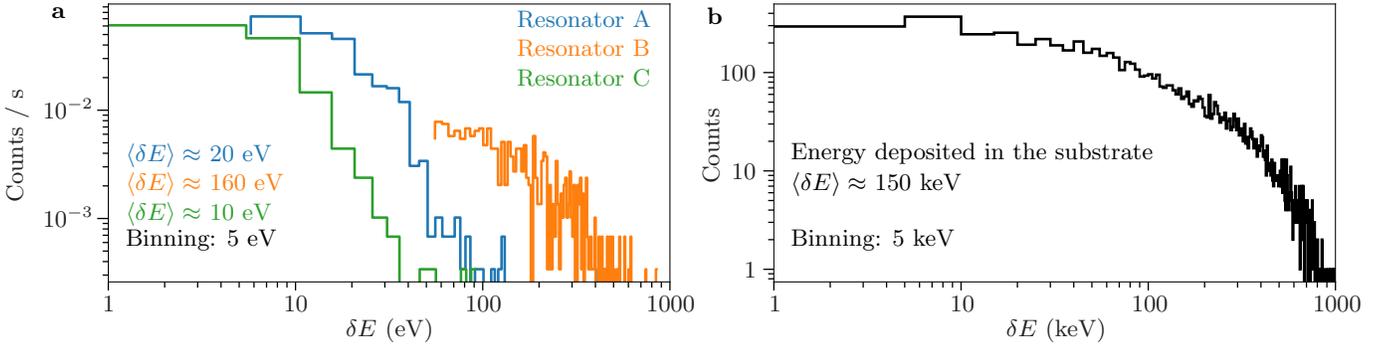
\caption{\textbf{Simulation of resonator efficiency under ThO$_2$ exposure.} \textbf{a}, distribution of the energy absorbed in the resonators in G under ThO$_2$ exposure. \textbf{b}, Monte Carlo simulation of the energy deposited by a ThO$_2$ source in the sapphire substrate.}
\label{A_fig_simu}
\end{center}
\end{figure*}

In Fig.~\ref{A_fig_simu}a we show the distribution of energy absorbed by the resonators in G with the Pb shield removed, and exposed to a ThO$_2$ source. We made a GEANT-4 \cite{geant4} based simulation to estimate the energy released in the substrate by the source. The simulation includes the sapphire chip, its copper holder, and the cryostat shields. The activity of the source was not precisely known, so we did not normalize the simulated events by time. We plot the results of the simulation in Fig.~\ref{A_fig_simu}b.
We can estimate the efficiency $\epsilon$ by dividing the average energy deposited in the resonator by the average energy deposited in the substrate. We obtain $\epsilon \sim  10^{-4}$,  $ 10^{-3}$, and $0.5 \times 10^{-4}$ for resonator A, B, and C, respectively.

This value could appear very small compared to values reported in Ref.~\cite{Moore:thesis, CARDANI2017338, Bellini:2016lgg}, therefore, to asses the phonon impedance mismatch at the substrate-film interface we performed a separate experiment in R. We measured a sample consisting of a 60~nm thick grAl kinetic inductance detector (KID) deposited on on a 2$\times$2~cm$^2$, 330~$\upmu$m thick sapphire substrate. The design was similar to the one described and depicted in Ref.~\cite{Bellini:2016lgg}, with an active surface of 2~mm$^2$. The chip was assembled in a copper holder hosting a $^{55}$Fe X-ray source, emitting typical X-rays at 5.9 and 6.4~keV. Due to their low energy, the X-rays are completely absorbed in the  substrate. This type of source has the disadvantage that, being permanently exposed to the sample, prevents the achievement of low-radioactivity levels. On the other hand, it produces events of well defined energy, in contrast to the broad spectrum emitted by the removable ThO$_2$ source (Fig.~\ref{A_fig_simu}b). This allows to obtain a precise measurement of the efficiency. Following the methods outlined in Refs.~\cite{Bellini:2016lgg, Moore2012}, we obtain the grAl KID efficiency of $0.32\%$.

We would like to emphasize that this value cannot be easily scaled to the A-C resonators discussed in the main text, which have 2-3 orders of magnitude smaller volume. Even if the efficiency scales almost linearly with the active volume of the KID~\cite{CARDANI2017338}, an extrapolation to a volume smaller by 2-3 orders of magnitude would not be reliable. Thus, we interpret the calibrated grAl KID efficiency of $3.2 \times 10^{-3}$ as an upper limit to the efficiency of the A-C resonators.

\vspace{10 mm}
\vspace{5 mm}

\end{document}

%% file: 1_intro.eps_tex
\begingroup%
  \makeatletter%
  \providecommand\color[2][]{%
    \errmessage{(Inkscape) Color is used for the text in Inkscape, but the package 'color.sty' is not loaded}%
    \renewcommand\color[2][]{}%
  }%
  \providecommand\transparent[1]{%
    \errmessage{(Inkscape) Transparency is used (non-zero) for the text in Inkscape, but the package 'transparent.sty' is not loaded}%
    \renewcommand\transparent[1]{}%
  }%
  \providecommand\rotatebox[2]{#2}%
  \newcommand*\fsize{\dimexpr\f@size pt\relax}%
  \newcommand*\lineheight[1]{\fontsize{\fsize}{#1\fsize}\selectfont}%
  \ifx\svgwidth\undefined%
    \setlength{\unitlength}{496.01449585bp}%
    \ifx\svgscale\undefined%
      \relax%
    \else%
      \setlength{\unitlength}{\unitlength * \real{\svgscale}}%
    \fi%
  \else%
    \setlength{\unitlength}{\svgwidth}%
  \fi%
  \global\let\svgwidth\undefined%
  \global\let\svgscale\undefined%
  \makeatother%
  \begin{picture}(1,0.53909714)%
    \lineheight{1}%
    \setlength\tabcolsep{0pt}%
    \put(0,0){\includegraphics[width=\unitlength]{1_intro.eps}}%
    \put(0.78976893,0.29912737){\color[rgb]{0.12156863,0.46666667,0.70588235}\makebox(0,0)[t]{\lineheight{1.25}\smash{\begin{tabular}[t]{c}0\end{tabular}}}}%
    \put(0.84922537,0.29912737){\color[rgb]{0.12156863,0.46666667,0.70588235}\makebox(0,0)[t]{\lineheight{1.25}\smash{\begin{tabular}[t]{c}100\end{tabular}}}}%
    \put(0.908682,0.29912737){\color[rgb]{0.12156863,0.46666667,0.70588235}\makebox(0,0)[t]{\lineheight{1.25}\smash{\begin{tabular}[t]{c}200\end{tabular}}}}%
    \put(0.85517407,0.27326767){\color[rgb]{0.12156863,0.46666667,0.70588235}\makebox(0,0)[t]{\lineheight{1.25}\smash{\begin{tabular}[t]{c}$-\delta f_0$, A (kHz)\end{tabular}}}}%
    \put(0.76550029,0.33391924){\color[rgb]{1,0.49803922,0.05490196}\makebox(0,0)[rt]{\lineheight{1.25}\smash{\begin{tabular}[t]{r}0\end{tabular}}}}%
    \put(0.76550029,0.36869706){\color[rgb]{1,0.49803922,0.05490196}\makebox(0,0)[rt]{\lineheight{1.25}\smash{\begin{tabular}[t]{r}50\end{tabular}}}}%
    \put(0.76550029,0.40347494){\color[rgb]{1,0.49803922,0.05490196}\makebox(0,0)[rt]{\lineheight{1.25}\smash{\begin{tabular}[t]{r}100\end{tabular}}}}%
    \put(0.76550029,0.43825274){\color[rgb]{1,0.49803922,0.05490196}\makebox(0,0)[rt]{\lineheight{1.25}\smash{\begin{tabular}[t]{r}150\end{tabular}}}}%
    \put(0.76550029,0.47303055){\color[rgb]{1,0.49803922,0.05490196}\makebox(0,0)[rt]{\lineheight{1.25}\smash{\begin{tabular}[t]{r}200\end{tabular}}}}%
    \put(0.94295605,0.33391924){\color[rgb]{0.17254902,0.62745098,0.17254902}\makebox(0,0)[lt]{\lineheight{1.25}\smash{\begin{tabular}[t]{l}0\end{tabular}}}}%
    \put(0.94295605,0.36869706){\color[rgb]{0.17254902,0.62745098,0.17254902}\makebox(0,0)[lt]{\lineheight{1.25}\smash{\begin{tabular}[t]{l}50\end{tabular}}}}%
    \put(0.94295605,0.40347494){\color[rgb]{0.17254902,0.62745098,0.17254902}\makebox(0,0)[lt]{\lineheight{1.25}\smash{\begin{tabular}[t]{l}100\end{tabular}}}}%
    \put(0.94295605,0.43825274){\color[rgb]{0.17254902,0.62745098,0.17254902}\makebox(0,0)[lt]{\lineheight{1.25}\smash{\begin{tabular}[t]{l}150\end{tabular}}}}%
    \put(0.94295605,0.47303055){\color[rgb]{0.17254902,0.62745098,0.17254902}\makebox(0,0)[lt]{\lineheight{1.25}\smash{\begin{tabular}[t]{l}200\end{tabular}}}}%
    \put(0.99153476,0.41286607){\color[rgb]{0.17254902,0.62745098,0.17254902}\rotatebox{90}{\makebox(0,0)[t]{\lineheight{1.25}\smash{\begin{tabular}[t]{c}$-\delta f_0$, C (kHz)\end{tabular}}}}}%
    \put(0.44677783,0.27242253){\makebox(0,0)[t]{\lineheight{1.25}\smash{\begin{tabular}[t]{c}t (s)\end{tabular}}}}%
    \put(0.2188382,0.41370894){\color[rgb]{0.12156863,0.46666667,0.70588235}\rotatebox{90}{\makebox(0,0)[t]{\lineheight{1.25}\smash{\begin{tabular}[t]{c}$\delta f_0$, A (kHz)\end{tabular}}}}}%
    \put(-0.6785863,-0.34791119){\color[rgb]{0,0,0}\makebox(0,0)[lt]{\begin{minipage}{0.20153546\unitlength}\raggedright \end{minipage}}}%
    \put(0.03061826,0.47030805){\color[rgb]{0.12156863,0.46666667,0.70588235}\makebox(0,0)[lt]{\begin{minipage}{0.03648302\unitlength}\raggedright A\end{minipage}}}%
    \put(0.03329351,0.29273424){\color[rgb]{1,0.49803922,0.05490196}\makebox(0,0)[lt]{\begin{minipage}{0.03648302\unitlength}\raggedright B\end{minipage}}}%
    \put(0.03240176,0.12264332){\color[rgb]{0.17254902,0.62745098,0.17254902}\makebox(0,0)[lt]{\begin{minipage}{0.03648302\unitlength}\raggedright C\end{minipage}}}%
    \put(-0.1370826,0.49593787){\color[rgb]{0.12156863,0.46666667,0.70588235}\makebox(0,0)[lt]{\begin{minipage}{0.50318875\unitlength}\centering $600\times10\;\upmu$m$^2$\end{minipage}}}%
    \put(-0.1370826,0.31965073){\color[rgb]{1,0.49803922,0.05490196}\makebox(0,0)[lt]{\begin{minipage}{0.50318875\unitlength}\centering $1000\times40\;\upmu$m$^2$\end{minipage}}}%
    \put(-0.1370826,0.15282218){\color[rgb]{0.17254902,0.62745098,0.17254902}\makebox(0,0)[lt]{\begin{minipage}{0.50318875\unitlength}\centering $420\times5\;\upmu$m$^2$\end{minipage}}}%
    \put(-0.13740551,0.44367352){\color[rgb]{0.12156863,0.46666667,0.70588235}\makebox(0,0)[lt]{\begin{minipage}{0.50318875\unitlength}\centering $f_0 \simeq 7.3$ GHz\end{minipage}}}%
    \put(-0.13740551,0.2671521){\color[rgb]{1,0.49803922,0.05490196}\makebox(0,0)[lt]{\begin{minipage}{0.50318875\unitlength}\centering $f_0 \simeq 7.4$ GHz\end{minipage}}}%
    \put(-0.13740551,0.09527769){\color[rgb]{0.17254902,0.62745098,0.17254902}\makebox(0,0)[lt]{\begin{minipage}{0.50318875\unitlength}\centering $f_0 \simeq 7.7$ GHz\end{minipage}}}%
    \put(-0.0059012,0.50619362){\color[rgb]{0,0,0}\makebox(0,0)[lt]{\begin{minipage}{0.22115396\unitlength}\raggedright \textbf{a}\end{minipage}}}%
    \put(0.15538828,-0.05075774){\color[rgb]{0,0,0}\rotatebox{90}{\makebox(0,0)[lt]{\begin{minipage}{0.50318873\unitlength}\centering 2 mm\end{minipage}}}}%
    \put(-0.00187483,0.02265801){\color[rgb]{0.57647059,0.6745098,0.65490196}\rotatebox{90}{\makebox(0,0)[lt]{\begin{minipage}{0.50318875\unitlength}\centering Full Al$_2$O$_3$ chip size: $15\times 8\; $mm$^2$ \end{minipage}}}}%
    \put(0.2251195,0.50619362){\color[rgb]{0,0,0}\makebox(0,0)[lt]{\begin{minipage}{0.22115395\unitlength}\raggedright \textbf{b}\end{minipage}}}%
    \put(0.2251195,0.27074077){\color[rgb]{0,0,0}\makebox(0,0)[lt]{\begin{minipage}{0.22115394\unitlength}\raggedright \textbf{d}\end{minipage}}}%
    \put(0.70387094,0.50619362){\color[rgb]{0,0,0}\makebox(0,0)[lt]{\begin{minipage}{0.22115395\unitlength}\raggedright \textbf{c}\end{minipage}}}%
    \put(0.61371524,0.01833381){\makebox(0,0)[t]{\lineheight{1.25}\smash{\begin{tabular}[t]{c}Energy absorbed in the resonator, $\delta E$ (eV)\end{tabular}}}}%
    \put(0.22037645,0.15943679){\rotatebox{90}{\makebox(0,0)[t]{\lineheight{1.25}\smash{\begin{tabular}[t]{c}Counts / s\end{tabular}}}}}%
    \put(0.72776275,0.41299023){\color[rgb]{1,0.49803922,0.05490196}\rotatebox{90}{\makebox(0,0)[t]{\lineheight{1.25}\smash{\begin{tabular}[t]{c}$-\delta f_0$, B (kHz)\end{tabular}}}}}%
    \put(0.7312363,0.24528876){\color[rgb]{0.12156863,0.46666667,0.70588235}\makebox(0,0)[lt]{\begin{minipage}{0.21754403\unitlength}\raggedright Resonator A\end{minipage}}}%
    \put(0.7312363,0.22813005){\color[rgb]{1,0.49803922,0.05490196}\makebox(0,0)[lt]{\begin{minipage}{0.21754403\unitlength}\raggedright Resonator B\end{minipage}}}%
    \put(0.7312363,0.21075866){\color[rgb]{0.17254902,0.62745098,0.17254902}\makebox(0,0)[lt]{\begin{minipage}{0.21754403\unitlength}\raggedright Resonator C\end{minipage}}}%
    \put(0.68042452,0.41294427){\color[rgb]{0.12156863,0.46666667,0.70588235}\rotatebox{90}{\makebox(0,0)[t]{\lineheight{1.25}\smash{\begin{tabular}[t]{c}$10^6\times \delta x_\mathrm{QP}$, A\end{tabular}}}}}%
    \put(0.27445215,0.29943005){\makebox(0,0)[t]{\lineheight{1.25}\smash{\begin{tabular}[t]{c}0.0\end{tabular}}}}%
    \put(0.32640281,0.29943002){\makebox(0,0)[t]{\lineheight{1.25}\smash{\begin{tabular}[t]{c}0.5\end{tabular}}}}%
    \put(0.37270513,0.29943002){\makebox(0,0)[t]{\lineheight{1.25}\smash{\begin{tabular}[t]{c}1.0\end{tabular}}}}%
    \put(0.41900742,0.29943002){\makebox(0,0)[t]{\lineheight{1.25}\smash{\begin{tabular}[t]{c}1.5\end{tabular}}}}%
    \put(0.46530971,0.29943002){\makebox(0,0)[t]{\lineheight{1.25}\smash{\begin{tabular}[t]{c}2.0\end{tabular}}}}%
    \put(0.5116122,0.29943002){\makebox(0,0)[t]{\lineheight{1.25}\smash{\begin{tabular}[t]{c}2.5\end{tabular}}}}%
    \put(0.55791443,0.29943002){\makebox(0,0)[t]{\lineheight{1.25}\smash{\begin{tabular}[t]{c}3.0\end{tabular}}}}%
    \put(0.60421672,0.29943002){\makebox(0,0)[t]{\lineheight{1.25}\smash{\begin{tabular}[t]{c}3.5\end{tabular}}}}%
    \put(0.27365324,0.33138935){\color[rgb]{0.12156863,0.46666667,0.70588235}\makebox(0,0)[rt]{\lineheight{1.25}\smash{\begin{tabular}[t]{r}-600\end{tabular}}}}%
    \put(0.27365324,0.38252993){\color[rgb]{0.12156863,0.46666667,0.70588235}\makebox(0,0)[rt]{\lineheight{1.25}\smash{\begin{tabular}[t]{r}-400\end{tabular}}}}%
    \put(0.27365324,0.43367039){\color[rgb]{0.12156863,0.46666667,0.70588235}\makebox(0,0)[rt]{\lineheight{1.25}\smash{\begin{tabular}[t]{r}-200\end{tabular}}}}%
    \put(0.27365324,0.48481097){\color[rgb]{0.12156863,0.46666667,0.70588235}\makebox(0,0)[rt]{\lineheight{1.25}\smash{\begin{tabular}[t]{r}0\end{tabular}}}}%
    \put(0.29097259,0.39192486){\color[rgb]{0,0,0}\makebox(0,0)[lt]{\begin{minipage}{0.15183515\unitlength}\raggedright $\delta x_\mathrm{QP} = -4 \frac{\delta f_0}{f_0}$\end{minipage}}}%
    \put(0.73123484,0.19331565){\color[rgb]{0,0,0}\makebox(0,0)[lt]{\begin{minipage}{0.47708621\unitlength}\raggedright Binning: 5 eV\end{minipage}}}%
    \put(0.62742168,0.34722369){\color[rgb]{0.12156863,0.46666667,0.70588235}\makebox(0,0)[lt]{\lineheight{1.25}\smash{\begin{tabular}[t]{l}300\end{tabular}}}}%
    \put(0.62742168,0.39308611){\color[rgb]{0.12156863,0.46666667,0.70588235}\makebox(0,0)[lt]{\lineheight{1.25}\smash{\begin{tabular}[t]{l}200\end{tabular}}}}%
    \put(0.62742168,0.43894857){\color[rgb]{0.12156863,0.46666667,0.70588235}\makebox(0,0)[lt]{\lineheight{1.25}\smash{\begin{tabular}[t]{l}100\end{tabular}}}}%
    \put(0.62742168,0.48481097){\color[rgb]{0.12156863,0.46666667,0.70588235}\makebox(0,0)[lt]{\lineheight{1.25}\smash{\begin{tabular}[t]{l}0\end{tabular}}}}%
    \put(0.31153985,0.04690331){\makebox(0,0)[t]{\lineheight{1.25}\smash{\begin{tabular}[t]{c}1\end{tabular}}}}%
    \put(0.51363087,0.04690331){\makebox(0,0)[t]{\lineheight{1.25}\smash{\begin{tabular}[t]{c}10\end{tabular}}}}%
    \put(0.71572195,0.04690331){\makebox(0,0)[t]{\lineheight{1.25}\smash{\begin{tabular}[t]{c}100\end{tabular}}}}%
    \put(0.917813,0.04690331){\makebox(0,0)[t]{\lineheight{1.25}\smash{\begin{tabular}[t]{c}1000\end{tabular}}}}%
    \put(0.27003152,0.1051488){\makebox(0,0)[rt]{\lineheight{1.25}\smash{\begin{tabular}[t]{r}$10^{-4}$\end{tabular}}}}%
    \put(0.27003152,0.15633622){\makebox(0,0)[rt]{\lineheight{1.25}\smash{\begin{tabular}[t]{r}$10^{-3}$\end{tabular}}}}%
    \put(0.27003152,0.20752359){\makebox(0,0)[rt]{\lineheight{1.25}\smash{\begin{tabular}[t]{r}$10^{-2}$\end{tabular}}}}%
    \put(0.29623681,0.17606061){\rotatebox{90}{\makebox(0,0)[t]{\lineheight{1.25}\smash{\begin{tabular}[t]{c}Counts / s\end{tabular}}}}}%
    \put(0.4520706,0.07523866){\makebox(0,0)[t]{\lineheight{1.25}\smash{\begin{tabular}[t]{c}\scriptsize $10^6\times \delta x_\mathrm{QP}$\end{tabular}}}}%
    \put(0.40054277,0.08644069){\makebox(0,0)[t]{\lineheight{1.25}\smash{\begin{tabular}[t]{c}\scriptsize 100\end{tabular}}}}%
    \put(0.50621767,0.08644069){\makebox(0,0)[t]{\lineheight{1.25}\smash{\begin{tabular}[t]{c}\scriptsize 1000\end{tabular}}}}%
    \put(0.3381104,0.16158248){\makebox(0,0)[rt]{\lineheight{1.25}\smash{\begin{tabular}[t]{r}\scriptsize$10^{-3}$\end{tabular}}}}%
    \put(0.3381104,0.19512105){\makebox(0,0)[rt]{\lineheight{1.25}\smash{\begin{tabular}[t]{r}\scriptsize$10^{-2}$\end{tabular}}}}%
    \put(0.3381104,0.2286596){\makebox(0,0)[rt]{\lineheight{1.25}\smash{\begin{tabular}[t]{r}\scriptsize $10^{-1}$\end{tabular}}}}%
    \put(0.34947959,0.14333046){\color[rgb]{0,0,0}\makebox(0,0)[lt]{\begin{minipage}{0.47708621\unitlength}\raggedright \scriptsize Binning: $5\times 10^{-5}$\end{minipage}}}%
    \put(0.3381104,0.12804391){\makebox(0,0)[rt]{\lineheight{1.25}\smash{\begin{tabular}[t]{r}\scriptsize$10^{-4}$\end{tabular}}}}%
  \end{picture}%
\endgroup%

%% file: 2_setups.eps_tex
\begingroup%
  \makeatletter%
  \providecommand\color[2][]{%
    \errmessage{(Inkscape) Color is used for the text in Inkscape, but the package 'color.sty' is not loaded}%
    \renewcommand\color[2][]{}%
  }%
  \providecommand\transparent[1]{%
    \errmessage{(Inkscape) Transparency is used (non-zero) for the text in Inkscape, but the package 'transparent.sty' is not loaded}%
    \renewcommand\transparent[1]{}%
  }%
  \providecommand\rotatebox[2]{#2}%
  \newcommand*\fsize{\dimexpr\f@size pt\relax}%
  \newcommand*\lineheight[1]{\fontsize{\fsize}{#1\fsize}\selectfont}%
  \ifx\svgwidth\undefined%
    \setlength{\unitlength}{810.97293884bp}%
    \ifx\svgscale\undefined%
      \relax%
    \else%
      \setlength{\unitlength}{\unitlength * \real{\svgscale}}%
    \fi%
  \else%
    \setlength{\unitlength}{\svgwidth}%
  \fi%
  \global\let\svgwidth\undefined%
  \global\let\svgscale\undefined%
  \makeatother%
  \begin{picture}(1,0.4435105)%
    \lineheight{1}%
    \setlength\tabcolsep{0pt}%
    \put(0,0){\includegraphics[width=\unitlength]{2_setups.eps}}%
    \put(0.1999194,0.04958668){\color[rgb]{0,0,0}\makebox(0,0)[lt]{\begin{minipage}{0.24589727\unitlength}\raggedright Stacked Pb bricks\end{minipage}}}%
    \put(0.05645302,0.04958668){\color[rgb]{0,0,0}\makebox(0,0)[lt]{\begin{minipage}{0.24589727\unitlength}\raggedright Granite\end{minipage}}}%
    \put(0.40307816,0.13608495){\color[rgb]{0,0,0}\makebox(0,0)[lt]{\begin{minipage}{0.24589727\unitlength}\raggedright Cu\end{minipage}}}%
    \put(0.20007599,0.09370856){\color[rgb]{0,0,0}\makebox(0,0)[lt]{\begin{minipage}{0.33096703\unitlength}\raggedright Cu (acid cleaned)\end{minipage}}}%
    \put(0.50007553,0.13583059){\color[rgb]{0,0,0}\makebox(0,0)[lt]{\begin{minipage}{0.27732776\unitlength}\raggedright Al\end{minipage}}}%
    \put(0.05602253,0.09431906){\color[rgb]{0,0,0}\makebox(0,0)[lt]{\begin{minipage}{0.27732776\unitlength}\raggedright $\upmu$-metal\end{minipage}}}%
    \put(0.01422248,0.13749778){\color[rgb]{0,0,0}\makebox(0,0)[lt]{\begin{minipage}{0.40301438\unitlength}\raggedright Glue ({\color{magenta}Ag paste} or {\color{blue}vacuum grease}) \end{minipage}}}%
    \put(0.07815762,0.43366238){\color[rgb]{0,0,0}\makebox(0,0)[lt]{\begin{minipage}{0.30020883\unitlength}\raggedright \textbf{K}arlsruhe ( {\color{magenta}07/2018} )\end{minipage}}}%
    \put(0.38354017,0.43419339){\color[rgb]{0,0,0}\makebox(0,0)[lt]{\begin{minipage}{0.23985482\unitlength}\raggedright \textbf{R}ome ( {\color{magenta}07/2019}, {\color{blue}11/2019} )\end{minipage}}}%
    \put(0.6973706,0.4356425){\color[rgb]{0,0,0}\makebox(0,0)[lt]{\begin{minipage}{0.23999205\unitlength}\centering \textbf{G}ran Sasso ( {\color{blue}04/2019} )\end{minipage}}}%
    \put(0.76108288,0.33526804){\color[rgb]{1,1,1}\rotatebox{-55.59719436}{\makebox(0,0)[lt]{\begin{minipage}{0.13150692\unitlength}\raggedright 1.4 km\end{minipage}}}}%
    \put(0.73356068,0.2607685){\color[rgb]{1,1,1}\makebox(0,0)[lt]{\begin{minipage}{0.3495616\unitlength}\raggedright highway A24 tunnel\end{minipage}}}%
    \put(0.82834334,0.29653916){\color[rgb]{0.66666667,0.87058824,0.52941176}\makebox(0,0)[lt]{\begin{minipage}{0.17014748\unitlength}\raggedright cryostat\end{minipage}}}%
    \put(0.15417748,0.37130244){\color[rgb]{0.16470588,0.83137255,1}\makebox(0,0)[lt]{\begin{minipage}{0.15002372\unitlength}\raggedright 25 mK\end{minipage}}}%
    \put(0.45025857,0.05136249){\color[rgb]{0,0,0}\makebox(0,0)[lt]{\begin{minipage}{0.31900103\unitlength}\raggedright ThO$_2$ source (on/off)\end{minipage}}}%
    \put(0.40307816,0.09431906){\color[rgb]{0,0,0}\makebox(0,0)[lt]{\begin{minipage}{0.40837691\unitlength}\raggedright Cryostat body (steel)\end{minipage}}}%
    \put(0.78505061,0.18172085){\color[rgb]{0.16470588,0.83137255,1}\makebox(0,0)[lt]{\begin{minipage}{0.15002372\unitlength}\raggedright 30 mK\end{minipage}}}%
    \put(0.4608702,0.36310144){\color[rgb]{0.16470588,0.83137255,1}\makebox(0,0)[lt]{\begin{minipage}{0.15002372\unitlength}\raggedright 20 mK\end{minipage}}}%
    \put(0.02914692,0.17313308){\color[rgb]{0,0,0}\makebox(0,0)[lt]{\begin{minipage}{0.48879046\unitlength}\raggedright grAl resonators \end{minipage}}}%
    \put(0.23805116,0.17352002){\color[rgb]{0,0,0}\makebox(0,0)[lt]{\begin{minipage}{0.48879046\unitlength}\raggedright Al$_2$O$_3$ substrate \end{minipage}}}%
    \put(0.55038932,0.17258516){\color[rgb]{0,0,0}\makebox(0,0)[lt]{\begin{minipage}{0.48879046\unitlength}\raggedright Al foil \end{minipage}}}%
    \put(0.53792146,0.38313767){\color[rgb]{0,0,0}\makebox(0,0)[lt]{\begin{minipage}{0.41220331\unitlength}\raggedright \scriptsize IN\end{minipage}}}%
    \put(0.43915519,0.38313767){\color[rgb]{0,0,0}\makebox(0,0)[lt]{\begin{minipage}{0.41220331\unitlength}\raggedright \scriptsize OUT\end{minipage}}}%
    \put(0.23744132,0.37166527){\color[rgb]{0,0,0}\makebox(0,0)[lt]{\begin{minipage}{0.21572257\unitlength}\raggedright \scriptsize -40 dB\end{minipage}}}%
    \put(0.11022889,0.39360742){\color[rgb]{0,0,0}\makebox(0,0)[lt]{\begin{minipage}{0.41220331\unitlength}\raggedright \scriptsize OUT\end{minipage}}}%
    \put(0.23697745,0.39360742){\color[rgb]{0,0,0}\makebox(0,0)[lt]{\begin{minipage}{0.41220331\unitlength}\raggedright \scriptsize IN\end{minipage}}}%
    \put(0.8526367,0.20405727){\color[rgb]{0,0,0}\makebox(0,0)[lt]{\begin{minipage}{0.41220331\unitlength}\raggedright \scriptsize IN\end{minipage}}}%
    \put(0.76744604,0.20405727){\color[rgb]{0,0,0}\makebox(0,0)[lt]{\begin{minipage}{0.41220331\unitlength}\raggedright \scriptsize OUT\end{minipage}}}%
    \put(0.23744132,0.34762006){\color[rgb]{0,0,0}\makebox(0,0)[lt]{\begin{minipage}{0.21572257\unitlength}\raggedright \scriptsize -20 dB\end{minipage}}}%
    \put(0.53352564,0.36212389){\color[rgb]{0,0,0}\makebox(0,0)[lt]{\begin{minipage}{0.21572257\unitlength}\raggedright \scriptsize -35 dB\end{minipage}}}%
    \put(0.53352564,0.33807868){\color[rgb]{0,0,0}\makebox(0,0)[lt]{\begin{minipage}{0.21572257\unitlength}\raggedright \scriptsize -30 dB\end{minipage}}}%
    \put(0.85364905,0.18182246){\color[rgb]{0,0,0}\makebox(0,0)[lt]{\begin{minipage}{0.21572257\unitlength}\raggedright \scriptsize -40 dB\end{minipage}}}%
    \put(0.85318195,0.1577667){\color[rgb]{0,0,0}\makebox(0,0)[lt]{\begin{minipage}{0.21572257\unitlength}\raggedright \scriptsize -30 dB\end{minipage}}}%
  \end{picture}%
\endgroup%

%% file: 3_rate_and_qi.eps_tex
\begingroup%
  \makeatletter%
  \providecommand\color[2][]{%
    \errmessage{(Inkscape) Color is used for the text in Inkscape, but the package 'color.sty' is not loaded}%
    \renewcommand\color[2][]{}%
  }%
  \providecommand\transparent[1]{%
    \errmessage{(Inkscape) Transparency is used (non-zero) for the text in Inkscape, but the package 'transparent.sty' is not loaded}%
    \renewcommand\transparent[1]{}%
  }%
  \providecommand\rotatebox[2]{#2}%
  \newcommand*\fsize{\dimexpr\f@size pt\relax}%
  \newcommand*\lineheight[1]{\fontsize{\fsize}{#1\fsize}\selectfont}%
  \ifx\svgwidth\undefined%
    \setlength{\unitlength}{260.3237915bp}%
    \ifx\svgscale\undefined%
      \relax%
    \else%
      \setlength{\unitlength}{\unitlength * \real{\svgscale}}%
    \fi%
  \else%
    \setlength{\unitlength}{\svgwidth}%
  \fi%
  \global\let\svgwidth\undefined%
  \global\let\svgscale\undefined%
  \makeatother%
  \begin{picture}(1,0.8891625)%
    \lineheight{1}%
    \setlength\tabcolsep{0pt}%
    \put(0,0){\includegraphics[width=\unitlength]{3_rate_and_qi.eps}}%
    \put(0.12515543,0.51166348){\makebox(0,0)[rt]{\lineheight{1.25}\smash{\begin{tabular}[t]{r}1\end{tabular}}}}%
    \put(0.12515543,0.62828422){\makebox(0,0)[rt]{\lineheight{1.25}\smash{\begin{tabular}[t]{r}10\end{tabular}}}}%
    \put(0.12515543,0.74490499){\makebox(0,0)[rt]{\lineheight{1.25}\smash{\begin{tabular}[t]{r}100\end{tabular}}}}%
    \put(0.05209558,0.66252521){\rotatebox{90}{\makebox(0,0)[t]{\lineheight{1.25}\smash{\begin{tabular}[t]{c}$\Gamma_B$ (mHz)\end{tabular}}}}}%
    \put(0.18621007,0.12845245){\makebox(0,0)[t]{\lineheight{1.25}\smash{\begin{tabular}[t]{c}K\end{tabular}}}}%
    \put(0.27758841,0.12845245){\makebox(0,0)[t]{\lineheight{1.25}\smash{\begin{tabular}[t]{c}R\end{tabular}}}}%
    \put(0.36896672,0.12845245){\makebox(0,0)[t]{\lineheight{1.25}\smash{\begin{tabular}[t]{c}G\end{tabular}}}}%
    \put(0.460345,0.12845245){\makebox(0,0)[t]{\lineheight{1.25}\smash{\begin{tabular}[t]{c}K\end{tabular}}}}%
    \put(0.55172337,0.12845245){\makebox(0,0)[t]{\lineheight{1.25}\smash{\begin{tabular}[t]{c}R\end{tabular}}}}%
    \put(0.64310168,0.12845245){\makebox(0,0)[t]{\lineheight{1.25}\smash{\begin{tabular}[t]{c}G\end{tabular}}}}%
    \put(0.73447999,0.12845245){\makebox(0,0)[t]{\lineheight{1.25}\smash{\begin{tabular}[t]{c}K\end{tabular}}}}%
    \put(0.82585831,0.12845245){\makebox(0,0)[t]{\lineheight{1.25}\smash{\begin{tabular}[t]{c}R\end{tabular}}}}%
    \put(0.91723668,0.12845245){\makebox(0,0)[t]{\lineheight{1.25}\smash{\begin{tabular}[t]{c}G\end{tabular}}}}%
    \put(0.07575542,0.16007413){\makebox(0,0)[lt]{\lineheight{1.25}\smash{\begin{tabular}[t]{l}10$^4$\end{tabular}}}}%
    \put(0.07575542,0.36576213){\makebox(0,0)[lt]{\lineheight{1.25}\smash{\begin{tabular}[t]{l}10$^5$\end{tabular}}}}%
    \put(0.05455404,0.3224813){\rotatebox{90}{\makebox(0,0)[t]{\lineheight{1.25}\smash{\begin{tabular}[t]{c}$Q_i \; @ \; \bar{n}=1$\end{tabular}}}}}%
    \put(0.14642144,0.06551331){\color[rgb]{1,0,1}\makebox(0,0)[lt]{\lineheight{1.25}\smash{\begin{tabular}[t]{l}silver paste\end{tabular}}}}%
    \put(0.34678469,0.06551331){\color[rgb]{0,0,1}\makebox(0,0)[lt]{\lineheight{1.25}\smash{\begin{tabular}[t]{l}vacuum grease\end{tabular}}}}%
    \put(0.66296579,0.06299863){\makebox(0,0)[lt]{\lineheight{1.25}\smash{\begin{tabular}[t]{l}lead off\end{tabular}}}}%
    \put(0.84498815,0.06640484){\makebox(0,0)[lt]{\lineheight{1.25}\smash{\begin{tabular}[t]{l}ThO$_2$\end{tabular}}}}%
    \put(0.27758841,0.83455263){\color[rgb]{0.12156863,0.46666667,0.70588235}\makebox(0,0)[t]{\lineheight{1.25}\smash{\begin{tabular}[t]{c}Resonator A\end{tabular}}}}%
    \put(0.55172337,0.83455263){\color[rgb]{1,0.49803922,0.05490196}\makebox(0,0)[t]{\lineheight{1.25}\smash{\begin{tabular}[t]{c}Resonator B\end{tabular}}}}%
    \put(0.82585831,0.83455263){\color[rgb]{0.17254902,0.62745098,0.17254902}\makebox(0,0)[t]{\lineheight{1.25}\smash{\begin{tabular}[t]{c}Resonator C\end{tabular}}}}%
  \end{picture}%
\endgroup%

%% file: A_schematics.eps_tex
\begingroup%
  \makeatletter%
  \providecommand\color[2][]{%
    \errmessage{(Inkscape) Color is used for the text in Inkscape, but the package 'color.sty' is not loaded}%
    \renewcommand\color[2][]{}%
  }%
  \providecommand\transparent[1]{%
    \errmessage{(Inkscape) Transparency is used (non-zero) for the text in Inkscape, but the package 'transparent.sty' is not loaded}%
    \renewcommand\transparent[1]{}%
  }%
  \providecommand\rotatebox[2]{#2}%
  \newcommand*\fsize{\dimexpr\f@size pt\relax}%
  \newcommand*\lineheight[1]{\fontsize{\fsize}{#1\fsize}\selectfont}%
  \ifx\svgwidth\undefined%
    \setlength{\unitlength}{553.47481795bp}%
    \ifx\svgscale\undefined%
      \relax%
    \else%
      \setlength{\unitlength}{\unitlength * \real{\svgscale}}%
    \fi%
  \else%
    \setlength{\unitlength}{\svgwidth}%
  \fi%
  \global\let\svgwidth\undefined%
  \global\let\svgscale\undefined%
  \makeatother%
  \begin{picture}(1,0.46264819)%
    \lineheight{1}%
    \setlength\tabcolsep{0pt}%
    \put(0,0){\includegraphics[width=\unitlength]{A_schematics.eps}}%
    \put(0.82237251,0.04234898){\color[rgb]{0,0,0}\makebox(0,0)[t]{\smash{\begin{tabular}[t]{c}Sample\\\end{tabular}}}}%
    \put(-0.1139373,0.33861691){\color[rgb]{0,0,0}\rotatebox{-90}{\makebox(0,0)[lt]{\begin{minipage}{0.07191375\unitlength}\centering \end{minipage}}}}%
    \put(-0.30770786,0.26618205){\color[rgb]{0,0,0}\rotatebox{-90}{\makebox(0,0)[lt]{\begin{minipage}{0.0832601\unitlength}\centering \end{minipage}}}}%
    \put(-0.09412211,0.64604578){\color[rgb]{0,0,0}\rotatebox{-90}{\makebox(0,0)[lt]{\begin{minipage}{0.06706396\unitlength}\centering \end{minipage}}}}%
    \put(0.28309782,0.29735453){\color[rgb]{0,0,0}\makebox(0,0)[t]{\smash{\begin{tabular}[t]{c}$-30$ dB\end{tabular}}}}%
    \put(0.28309782,0.24256614){\color[rgb]{0,0,0}\makebox(0,0)[t]{\smash{\begin{tabular}[t]{c}$-10$ dB\end{tabular}}}}%
    \put(0.10082576,0.11153875){\color[rgb]{0,0,0}\makebox(0,0)[t]{\smash{\begin{tabular}[t]{c}IR \\Filter\end{tabular}}}}%
    \put(0.18820256,0.04234898){\color[rgb]{0,0,0}\makebox(0,0)[t]{\smash{\begin{tabular}[t]{c}Sample\\\end{tabular}}}}%
    \put(0.08660681,0.17777809){\color[rgb]{0,0,0}\makebox(0,0)[t]{\smash{\begin{tabular}[t]{c}$12$ GHz\end{tabular}}}}%
    \put(0.08676261,0.30052043){\color[rgb]{0,0,0}\makebox(0,0)[t]{\smash{\begin{tabular}[t]{c}$+40$ dB\end{tabular}}}}%
    \put(0.00030594,0.12715946){\color[rgb]{0,0,0}\rotatebox{-90}{\makebox(0,0)[lt]{\begin{minipage}{0.2954348\unitlength}\centering \end{minipage}}}}%
    \put(0.28309782,0.14497247){\color[rgb]{0,0,0}\makebox(0,0)[t]{\smash{\begin{tabular}[t]{c}$-20$ dB\end{tabular}}}}%
    \put(0.18606644,0.32868691){\color[rgb]{0,0.2,0.50196078}\makebox(0,0)[t]{\smash{\begin{tabular}[t]{c}$4$ K\end{tabular}}}}%
    \put(0.18622153,0.27269903){\color[rgb]{0,0.33333333,0.83137255}\makebox(0,0)[t]{\smash{\begin{tabular}[t]{c}$500$ mk \end{tabular}}}}%
    \put(0.18625107,0.21198018){\color[rgb]{0,0.4,1}\makebox(0,0)[t]{\smash{\begin{tabular}[t]{c}$25$ mk\end{tabular}}}}%
    \put(0.28294202,0.17744849){\color[rgb]{0,0,0}\makebox(0,0)[t]{\smash{\begin{tabular}[t]{c}$12$ GHz\end{tabular}}}}%
    \put(0.27352958,0.11153875){\color[rgb]{0,0,0}\makebox(0,0)[t]{\smash{\begin{tabular}[t]{c}IR \\Filter\end{tabular}}}}%
    \put(0.0867626,0.36412856){\color[rgb]{0,0,0}\makebox(0,0)[t]{\smash{\begin{tabular}[t]{c}$+40$ dB\end{tabular}}}}%
    \put(0.18726498,0.38895054){\color[rgb]{0,0,0}\makebox(0,0)[t]{\smash{\begin{tabular}[t]{c}VNA\end{tabular}}}}%
    \put(0.18793144,0.43022581){\color[rgb]{0,0,0}\makebox(0,0)[t]{\smash{\begin{tabular}[t]{c}\textbf{K}\end{tabular}}}}%
    \put(-0.08764817,0.58489855){\color[rgb]{0,0,0}\makebox(0,0)[lt]{\begin{minipage}{0.48191783\unitlength}\centering \end{minipage}}}%
    \put(0.50297673,0.4316089){\color[rgb]{0,0,0}\makebox(0,0)[t]{\smash{\begin{tabular}[t]{c}\textbf{R}\end{tabular}}}}%
    \put(0.82227026,0.4316089){\color[rgb]{0,0,0}\makebox(0,0)[t]{\smash{\begin{tabular}[t]{c}\textbf{G}\end{tabular}}}}%
    \put(0.59941128,0.25715797){\color[rgb]{0,0,0}\makebox(0,0)[t]{\smash{\begin{tabular}[t]{c}$-35$ dB\end{tabular}}}}%
    \put(0.57459018,0.29540237){\color[rgb]{1,1,1}\rotatebox{-90}{\makebox(0,0)[t]{\smash{\begin{tabular}[t]{c}-30dB\end{tabular}}}}}%
    \put(0.50451602,0.04234898){\color[rgb]{0,0,0}\makebox(0,0)[t]{\smash{\begin{tabular}[t]{c}Sample\\\end{tabular}}}}%
    \put(0.40307607,0.25715797){\color[rgb]{0,0,0}\makebox(0,0)[t]{\smash{\begin{tabular}[t]{c}$+40$ dB\end{tabular}}}}%
    \put(0.59941128,0.11394981){\color[rgb]{0,0,0}\makebox(0,0)[t]{\smash{\begin{tabular}[t]{c}$-30$ dB\end{tabular}}}}%
    \put(0.5023799,0.28803461){\color[rgb]{0,0.2,0.50196078}\makebox(0,0)[t]{\smash{\begin{tabular}[t]{c}$4$ K\end{tabular}}}}%
    \put(0.50256453,0.14422635){\color[rgb]{0,0.4,1}\makebox(0,0)[t]{\smash{\begin{tabular}[t]{c}$20$ mk\end{tabular}}}}%
    \put(0.40307607,0.36412856){\color[rgb]{0,0,0}\makebox(0,0)[t]{\smash{\begin{tabular}[t]{c}$+40$ dB\end{tabular}}}}%
    \put(0.50357844,0.38895054){\color[rgb]{0,0,0}\makebox(0,0)[t]{\smash{\begin{tabular}[t]{c}VNA\end{tabular}}}}%
    \put(0.59941129,0.36494378){\color[rgb]{0,0,0}\makebox(0,0)[t]{\smash{\begin{tabular}[t]{c}$-30$ dB\end{tabular}}}}%
    \put(0.91726495,0.25715797){\color[rgb]{0,0,0}\makebox(0,0)[t]{\smash{\begin{tabular}[t]{c}$-40$ dB\end{tabular}}}}%
    \put(0.89244456,0.29540237){\color[rgb]{1,1,1}\rotatebox{-90}{\makebox(0,0)[t]{\smash{\begin{tabular}[t]{c}-30dB\end{tabular}}}}}%
    \put(0.72093544,0.25715797){\color[rgb]{0,0,0}\makebox(0,0)[t]{\smash{\begin{tabular}[t]{c}$+40$ dB\end{tabular}}}}%
    \put(0.91726495,0.11394981){\color[rgb]{0,0,0}\makebox(0,0)[t]{\smash{\begin{tabular}[t]{c}$-30$ dB\end{tabular}}}}%
    \put(0.82023639,0.28803461){\color[rgb]{0,0.2,0.50196078}\makebox(0,0)[t]{\smash{\begin{tabular}[t]{c}$4$ K\end{tabular}}}}%
    \put(0.82042109,0.14422635){\color[rgb]{0,0.4,1}\makebox(0,0)[t]{\smash{\begin{tabular}[t]{c}$30$ mk\end{tabular}}}}%
    \put(0.72093544,0.36412856){\color[rgb]{0,0,0}\makebox(0,0)[t]{\smash{\begin{tabular}[t]{c}$+40$ dB\end{tabular}}}}%
    \put(0.82143493,0.38895054){\color[rgb]{0,0,0}\makebox(0,0)[t]{\smash{\begin{tabular}[t]{c}VNA\end{tabular}}}}%
  \end{picture}%
\endgroup%

%% file: A_cryostat.eps_tex
\begingroup%
  \makeatletter%
  \providecommand\color[2][]{%
    \errmessage{(Inkscape) Color is used for the text in Inkscape, but the package 'color.sty' is not loaded}%
    \renewcommand\color[2][]{}%
  }%
  \providecommand\transparent[1]{%
    \errmessage{(Inkscape) Transparency is used (non-zero) for the text in Inkscape, but the package 'transparent.sty' is not loaded}%
    \renewcommand\transparent[1]{}%
  }%
  \providecommand\rotatebox[2]{#2}%
  \newcommand*\fsize{\dimexpr\f@size pt\relax}%
  \newcommand*\lineheight[1]{\fontsize{\fsize}{#1\fsize}\selectfont}%
  \ifx\svgwidth\undefined%
    \setlength{\unitlength}{529.89664393bp}%
    \ifx\svgscale\undefined%
      \relax%
    \else%
      \setlength{\unitlength}{\unitlength * \real{\svgscale}}%
    \fi%
  \else%
    \setlength{\unitlength}{\svgwidth}%
  \fi%
  \global\let\svgwidth\undefined%
  \global\let\svgscale\undefined%
  \makeatother%
  \begin{picture}(1,0.51316927)%
    \lineheight{1}%
    \setlength\tabcolsep{0pt}%
    \put(0,0){\includegraphics[width=\unitlength]{A_cryostat.eps}}%
    \put(0.91687159,0.02970428){\color[rgb]{1,1,1}\makebox(0,0)[lt]{\begin{minipage}{0.09793721\unitlength}\centering 20 cm\end{minipage}}}%
    \put(0.64479592,0.44360937){\color[rgb]{1,1,1}\makebox(0,0)[lt]{\begin{minipage}{0.27296433\unitlength}\centering $\upmu$-metal shield\end{minipage}}}%
    \put(0.71325163,0.18521321){\color[rgb]{1,1,1}\makebox(0,0)[lt]{\begin{minipage}{0.27296433\unitlength}\centering Pb bricks\end{minipage}}}%
    \put(0.38021353,0.02970428){\color[rgb]{1,1,1}\makebox(0,0)[lt]{\begin{minipage}{0.09793721\unitlength}\centering 2 cm\end{minipage}}}%
    \put(0.02016059,0.0298038){\color[rgb]{0,0,0}\makebox(0,0)[lt]{\begin{minipage}{0.09793721\unitlength}\centering 1 cm\end{minipage}}}%
    \put(0.07975884,0.21396612){\color[rgb]{1,1,1}\makebox(0,0)[lt]{\begin{minipage}{0.30239002\unitlength}\centering vacuum grease\end{minipage}}}%
    \put(0.01333803,0.50462836){\color[rgb]{0,0,0}\makebox(0,0)[lt]{\begin{minipage}{0.10026441\unitlength}\raggedright \textbf{a}\end{minipage}}}%
    \put(0.37033619,0.50462836){\color[rgb]{0,0,0}\makebox(0,0)[lt]{\begin{minipage}{0.14509518\unitlength}\raggedright \textbf{b}\end{minipage}}}%
    \put(0.63177528,0.50462836){\color[rgb]{0,0,0}\makebox(0,0)[lt]{\begin{minipage}{0.11172107\unitlength}\raggedright \textbf{c}\end{minipage}}}%
    \put(0.40351846,0.42071428){\color[rgb]{1,1,1}\makebox(0,0)[lt]{\begin{minipage}{0.27296433\unitlength}\centering barrel lid\end{minipage}}}%
  \end{picture}%
\endgroup%

%% file: A_drift.eps_tex
\begingroup%
  \makeatletter%
  \providecommand\color[2][]{%
    \errmessage{(Inkscape) Color is used for the text in Inkscape, but the package 'color.sty' is not loaded}%
    \renewcommand\color[2][]{}%
  }%
  \providecommand\transparent[1]{%
    \errmessage{(Inkscape) Transparency is used (non-zero) for the text in Inkscape, but the package 'transparent.sty' is not loaded}%
    \renewcommand\transparent[1]{}%
  }%
  \providecommand\rotatebox[2]{#2}%
  \newcommand*\fsize{\dimexpr\f@size pt\relax}%
  \newcommand*\lineheight[1]{\fontsize{\fsize}{#1\fsize}\selectfont}%
  \ifx\svgwidth\undefined%
    \setlength{\unitlength}{518.40002441bp}%
    \ifx\svgscale\undefined%
      \relax%
    \else%
      \setlength{\unitlength}{\unitlength * \real{\svgscale}}%
    \fi%
  \else%
    \setlength{\unitlength}{\svgwidth}%
  \fi%
  \global\let\svgwidth\undefined%
  \global\let\svgscale\undefined%
  \makeatother%
  \begin{picture}(1,0.33333332)%
    \lineheight{1}%
    \setlength\tabcolsep{0pt}%
    \put(0,0){\includegraphics[width=\unitlength]{A_drift.eps}}%
    \put(0.55849522,0.03564663){\makebox(0,0)[t]{\lineheight{1.25}\smash{\begin{tabular}[t]{c}0\end{tabular}}}}%
    \put(0.61605488,0.03564663){\makebox(0,0)[t]{\lineheight{1.25}\smash{\begin{tabular}[t]{c}10\end{tabular}}}}%
    \put(0.67361454,0.03564663){\makebox(0,0)[t]{\lineheight{1.25}\smash{\begin{tabular}[t]{c}20\end{tabular}}}}%
    \put(0.7311742,0.03564663){\makebox(0,0)[t]{\lineheight{1.25}\smash{\begin{tabular}[t]{c}30\end{tabular}}}}%
    \put(0.7887338,0.03564663){\makebox(0,0)[t]{\lineheight{1.25}\smash{\begin{tabular}[t]{c}40\end{tabular}}}}%
    \put(0.84629346,0.03564663){\makebox(0,0)[t]{\lineheight{1.25}\smash{\begin{tabular}[t]{c}50\end{tabular}}}}%
    \put(0.90385312,0.03564663){\makebox(0,0)[t]{\lineheight{1.25}\smash{\begin{tabular}[t]{c}60\end{tabular}}}}%
    \put(0.96141278,0.03564663){\makebox(0,0)[t]{\lineheight{1.25}\smash{\begin{tabular}[t]{c}70\end{tabular}}}}%
    \put(0.75995397,0.01709528){\makebox(0,0)[t]{\lineheight{1.25}\smash{\begin{tabular}[t]{c}Weeks\end{tabular}}}}%
    \put(0.52773981,0.0624804){\makebox(0,0)[rt]{\lineheight{1.25}\smash{\begin{tabular}[t]{r}−100\end{tabular}}}}%
    \put(0.52773981,0.10968219){\makebox(0,0)[rt]{\lineheight{1.25}\smash{\begin{tabular}[t]{r}−80\end{tabular}}}}%
    \put(0.52773981,0.15688397){\makebox(0,0)[rt]{\lineheight{1.25}\smash{\begin{tabular}[t]{r}−60\end{tabular}}}}%
    \put(0.52773981,0.20408578){\makebox(0,0)[rt]{\lineheight{1.25}\smash{\begin{tabular}[t]{r}−40\end{tabular}}}}%
    \put(0.52773981,0.25128756){\makebox(0,0)[rt]{\lineheight{1.25}\smash{\begin{tabular}[t]{r}−20\end{tabular}}}}%
    \put(0.52773981,0.29848935){\makebox(0,0)[rt]{\lineheight{1.25}\smash{\begin{tabular}[t]{r}0\end{tabular}}}}%
    \put(0.48799673,0.18518517){\rotatebox{90}{\makebox(0,0)[t]{\lineheight{1.25}\smash{\begin{tabular}[t]{c}Frequency drift (MHz)\end{tabular}}}}}%
    \put(0.03437759,0.26100046){\color[rgb]{0.19607843,0.51372549,0.72941176}\makebox(0,0)[lt]{\begin{minipage}{0.27693005\unitlength}\centering Resonator A\end{minipage}}}%
    \put(0.14931046,0.26100046){\color[rgb]{1,0.49803922,0.05490196}\makebox(0,0)[lt]{\begin{minipage}{0.27693005\unitlength}\centering Resonator B\end{minipage}}}%
    \put(0.26364429,0.26089874){\color[rgb]{0.21176471,0.61960784,0.17647059}\makebox(0,0)[lt]{\begin{minipage}{0.27693005\unitlength}\centering Resonator C\end{minipage}}}%
    \put(-0.07162248,0.1984232){\color[rgb]{0,0,0}\makebox(0,0)[lt]{\begin{minipage}{0.25702004\unitlength}\centering K\end{minipage}}}%
    \put(-0.07095185,0.15831653){\color[rgb]{0,0,0}\makebox(0,0)[lt]{\begin{minipage}{0.25702004\unitlength}\centering G\end{minipage}}}%
    \put(-0.08822254,0.11956998){\color[rgb]{0,0,0}\makebox(0,0)[lt]{\begin{minipage}{0.25702004\unitlength}\centering R, vg\end{minipage}}}%
    \put(-0.08768,0.07945576){\color[rgb]{0,0,0}\makebox(0,0)[lt]{\begin{minipage}{0.25702004\unitlength}\centering R, sp\end{minipage}}}%
    \put(0.17343777,0.18477723){\color[rgb]{0,0,0}\makebox(0,0)[t]{\lineheight{1.25}\smash{\begin{tabular}[t]{c}7.293158\end{tabular}}}}%
    \put(0.17334358,0.14466302){\color[rgb]{0,0,0}\makebox(0,0)[t]{\lineheight{1.25}\smash{\begin{tabular}[t]{c}7.243984\end{tabular}}}}%
    \put(0.17368266,0.10454883){\color[rgb]{0,0,0}\makebox(0,0)[t]{\lineheight{1.25}\smash{\begin{tabular}[t]{c}7.239192\end{tabular}}}}%
    \put(0.17334358,0.06443461){\color[rgb]{0,0,0}\makebox(0,0)[t]{\lineheight{1.25}\smash{\begin{tabular}[t]{c}7.193234\end{tabular}}}}%
    \put(0.28787333,0.18477723){\color[rgb]{0,0,0}\makebox(0,0)[t]{\lineheight{1.25}\smash{\begin{tabular}[t]{c}7.465360\end{tabular}}}}%
    \put(0.28779421,0.14466302){\color[rgb]{0,0,0}\makebox(0,0)[t]{\lineheight{1.25}\smash{\begin{tabular}[t]{c}7.419364\end{tabular}}}}%
    \put(0.28802026,0.10454883){\color[rgb]{0,0,0}\makebox(0,0)[t]{\lineheight{1.25}\smash{\begin{tabular}[t]{c}7.415607\end{tabular}}}}%
    \put(0.28813329,0.06443461){\color[rgb]{0,0,0}\makebox(0,0)[t]{\lineheight{1.25}\smash{\begin{tabular}[t]{c}7.369042\end{tabular}}}}%
    \put(0.40224481,0.18477723){\color[rgb]{0,0,0}\makebox(0,0)[t]{\lineheight{1.25}\smash{\begin{tabular}[t]{c}7.709244\end{tabular}}}}%
    \put(0.40232393,0.14466302){\color[rgb]{0,0,0}\makebox(0,0)[t]{\lineheight{1.25}\smash{\begin{tabular}[t]{c}7.658760\end{tabular}}}}%
    \put(0.40247086,0.10454883){\color[rgb]{0,0,0}\makebox(0,0)[t]{\lineheight{1.25}\smash{\begin{tabular}[t]{c}7.654537\end{tabular}}}}%
    \put(0.40242942,0.06443461){\color[rgb]{0,0,0}\makebox(0,0)[t]{\lineheight{1.25}\smash{\begin{tabular}[t]{c}7.610113\end{tabular}}}}%
    \put(0.1168942,0.3070316){\color[rgb]{0,0,0}\makebox(0,0)[lt]{\begin{minipage}{0.35750852\unitlength}\centering \textbf{Resonant frequency (GHz)}\end{minipage}}}%
    \put(-0.11458728,0.30716724){\color[rgb]{0,0,0}\makebox(0,0)[lt]{\begin{minipage}{0.35750852\unitlength}\centering \textbf{Setup}\end{minipage}}}%
    \put(0.520412,0.31126329){\color[rgb]{0,0,0}\makebox(0,0)[lt]{\begin{minipage}{0.25702004\unitlength}\centering K (07/2018)\end{minipage}}}%
    \put(0.59275193,0.18294313){\color[rgb]{0,0,0}\makebox(0,0)[lt]{\begin{minipage}{0.25702004\unitlength}\centering G (04/2019)\end{minipage}}}%
    \put(0.76677544,0.21274081){\color[rgb]{0,0,0}\makebox(0,0)[lt]{\begin{minipage}{0.25702004\unitlength}\centering R, vg (07/2019)\end{minipage}}}%
    \put(0.75194631,0.07922139){\color[rgb]{0,0,0}\makebox(0,0)[lt]{\begin{minipage}{0.25702004\unitlength}\centering R, sp (11/2019)\end{minipage}}}%
    \put(0.67832764,0.30546076){\color[rgb]{1,0,0}\makebox(0,0)[lt]{\begin{minipage}{0.45221844\unitlength}\centering 1.3 MHz / week\end{minipage}}}%
  \end{picture}%
\endgroup%

%% file: A_corr.eps_tex
\begingroup%
  \makeatletter%
  \providecommand\color[2][]{%
    \errmessage{(Inkscape) Color is used for the text in Inkscape, but the package 'color.sty' is not loaded}%
    \renewcommand\color[2][]{}%
  }%
  \providecommand\transparent[1]{%
    \errmessage{(Inkscape) Transparency is used (non-zero) for the text in Inkscape, but the package 'transparent.sty' is not loaded}%
    \renewcommand\transparent[1]{}%
  }%
  \providecommand\rotatebox[2]{#2}%
  \newcommand*\fsize{\dimexpr\f@size pt\relax}%
  \newcommand*\lineheight[1]{\fontsize{\fsize}{#1\fsize}\selectfont}%
  \ifx\svgwidth\undefined%
    \setlength{\unitlength}{524.38928223bp}%
    \ifx\svgscale\undefined%
      \relax%
    \else%
      \setlength{\unitlength}{\unitlength * \real{\svgscale}}%
    \fi%
  \else%
    \setlength{\unitlength}{\svgwidth}%
  \fi%
  \global\let\svgwidth\undefined%
  \global\let\svgscale\undefined%
  \makeatother%
  \begin{picture}(1,0.41381471)%
    \lineheight{1}%
    \setlength\tabcolsep{0pt}%
    \put(0,0){\includegraphics[width=\unitlength]{A_corr.eps}}%
    \put(0.22115722,0.12521595){\color[rgb]{0.11764706,0.56470588,1}\makebox(0,0)[t]{\lineheight{1.25}\smash{\begin{tabular}[t]{c}$f_A$\end{tabular}}}}%
    \put(0.27443587,0.08763814){\color[rgb]{0.11764706,0.56470588,1}\makebox(0,0)[t]{\lineheight{1.25}\smash{\begin{tabular}[t]{c}$f_\mathrm{A} = f_{0, \mathrm{A @ G}} - 285$ kHz\end{tabular}}}}%
    \put(0.31509841,0.12521593){\color[rgb]{0.60392157,0.80392157,0.19607843}\makebox(0,0)[t]{\lineheight{1.25}\smash{\begin{tabular}[t]{c}$f_C$\end{tabular}}}}%
    \put(0.07481829,0.19966062){\makebox(0,0)[rt]{\lineheight{1.25}\smash{\begin{tabular}[t]{r}-0.5\end{tabular}}}}%
    \put(0.07481829,0.25932668){\makebox(0,0)[rt]{\lineheight{1.25}\smash{\begin{tabular}[t]{r}0.0\end{tabular}}}}%
    \put(0.07481829,0.31899274){\makebox(0,0)[rt]{\lineheight{1.25}\smash{\begin{tabular}[t]{r}0.5\end{tabular}}}}%
    \put(0.07481829,0.37865879){\makebox(0,0)[rt]{\lineheight{1.25}\smash{\begin{tabular}[t]{r}1.0\end{tabular}}}}%
    \put(0.0349327,0.27411877){\rotatebox{90}{\makebox(0,0)[t]{\lineheight{1.25}\smash{\begin{tabular}[t]{c}arg\{S$_{11}$\} (rads)\end{tabular}}}}}%
    \put(-0.1051525,0.39611957){\color[rgb]{0,0,0}\makebox(0,0)[lt]{\begin{minipage}{0.22774537\unitlength}\centering \textbf{a}\end{minipage}}}%
    \put(0.36773688,0.39795373){\color[rgb]{0,0,0}\makebox(0,0)[lt]{\begin{minipage}{0.2334129\unitlength}\centering \textbf{c}\end{minipage}}}%
    \put(0.5598483,0.04000693){\makebox(0,0)[t]{\lineheight{1.25}\smash{\begin{tabular}[t]{c}0\end{tabular}}}}%
    \put(0.60749726,0.04000693){\makebox(0,0)[t]{\lineheight{1.25}\smash{\begin{tabular}[t]{c}200\end{tabular}}}}%
    \put(0.65514616,0.04000693){\makebox(0,0)[t]{\lineheight{1.25}\smash{\begin{tabular}[t]{c}400\end{tabular}}}}%
    \put(0.70279506,0.04000693){\makebox(0,0)[t]{\lineheight{1.25}\smash{\begin{tabular}[t]{c}600\end{tabular}}}}%
    \put(0.75044396,0.04000693){\makebox(0,0)[t]{\lineheight{1.25}\smash{\begin{tabular}[t]{c}800\end{tabular}}}}%
    \put(0.79809292,0.04000693){\makebox(0,0)[t]{\lineheight{1.25}\smash{\begin{tabular}[t]{c}1000\end{tabular}}}}%
    \put(0.84574182,0.04000693){\makebox(0,0)[t]{\lineheight{1.25}\smash{\begin{tabular}[t]{c}1200\end{tabular}}}}%
    \put(0.89339079,0.04000693){\makebox(0,0)[t]{\lineheight{1.25}\smash{\begin{tabular}[t]{c}1400\end{tabular}}}}%
    \put(0.738496,0.02158084){\makebox(0,0)[t]{\lineheight{1.25}\smash{\begin{tabular}[t]{c}time (s)\end{tabular}}}}%
    \put(0.53632617,0.05603906){\color[rgb]{0.11764706,0.56470588,1}\makebox(0,0)[rt]{\lineheight{1.25}\smash{\begin{tabular}[t]{r}-0.5\end{tabular}}}}%
    \put(0.53632617,0.11947703){\color[rgb]{0.11764706,0.56470588,1}\makebox(0,0)[rt]{\lineheight{1.25}\smash{\begin{tabular}[t]{r}-0.4\end{tabular}}}}%
    \put(0.53632617,0.18390026){\color[rgb]{0.11764706,0.56470588,1}\makebox(0,0)[rt]{\lineheight{1.25}\smash{\begin{tabular}[t]{r}-0.3\end{tabular}}}}%
    \put(0.53632617,0.24783085){\color[rgb]{0.11764706,0.56470588,1}\makebox(0,0)[rt]{\lineheight{1.25}\smash{\begin{tabular}[t]{r}-0.2\end{tabular}}}}%
    \put(0.53632617,0.31126884){\color[rgb]{0.11764706,0.56470588,1}\makebox(0,0)[rt]{\lineheight{1.25}\smash{\begin{tabular}[t]{r}-0.1\end{tabular}}}}%
    \put(0.53632617,0.37470682){\color[rgb]{0.11764706,0.56470588,1}\makebox(0,0)[rt]{\lineheight{1.25}\smash{\begin{tabular}[t]{r}0.0\end{tabular}}}}%
    \put(0.49555382,0.22902833){\color[rgb]{0.11764706,0.56470588,1}\rotatebox{90}{\makebox(0,0)[t]{\lineheight{1.25}\smash{\begin{tabular}[t]{c}$\varphi - \varphi(t=0) \; @ \; f_A$\end{tabular}}}}}%
    \put(0.94066595,0.08998206){\color[rgb]{0.60392157,0.80392157,0.19607843}\makebox(0,0)[lt]{\lineheight{1.25}\smash{\begin{tabular}[t]{l}0.0\end{tabular}}}}%
    \put(0.94066595,0.15774854){\color[rgb]{0.60392157,0.80392157,0.19607843}\makebox(0,0)[lt]{\lineheight{1.25}\smash{\begin{tabular}[t]{l}-0.2\end{tabular}}}}%
    \put(0.94066595,0.22534074){\color[rgb]{0.60392157,0.80392157,0.19607843}\makebox(0,0)[lt]{\lineheight{1.25}\smash{\begin{tabular}[t]{l}-0.4\end{tabular}}}}%
    \put(0.94066595,0.29310716){\color[rgb]{0.60392157,0.80392157,0.19607843}\makebox(0,0)[lt]{\lineheight{1.25}\smash{\begin{tabular}[t]{l}-0.6\end{tabular}}}}%
    \put(0.94066595,0.36122197){\color[rgb]{0.60392157,0.80392157,0.19607843}\makebox(0,0)[lt]{\lineheight{1.25}\smash{\begin{tabular}[t]{l}-0.8\end{tabular}}}}%
    \put(0.98957149,0.22902833){\color[rgb]{0.60392157,0.80392157,0.19607843}\rotatebox{90}{\makebox(0,0)[t]{\lineheight{1.25}\smash{\begin{tabular}[t]{c}$\varphi - \varphi(t=0) \; @ \; f_C$\end{tabular}}}}}%
    \put(0.28238148,0.38445455){\color[rgb]{0.11764706,0.56470588,1}\makebox(0,0)[lt]{\begin{minipage}{0.20524207\unitlength}\centering Resonator A\end{minipage}}}%
    \put(0.28249777,0.3479679){\color[rgb]{0.60392157,0.80392157,0.19607843}\makebox(0,0)[lt]{\begin{minipage}{0.20524207\unitlength}\centering Resonator C\end{minipage}}}%
    \put(0.27443587,0.06475445){\color[rgb]{0.60392157,0.80392157,0.19607843}\makebox(0,0)[t]{\lineheight{1.25}\smash{\begin{tabular}[t]{c}$f_\mathrm{C} = f_{0, \mathrm{C @ G}} - 260$ kHz\end{tabular}}}}%
  \end{picture}%
\endgroup%

%% file: A_qi.eps_tex
\begingroup%
  \makeatletter%
  \providecommand\color[2][]{%
    \errmessage{(Inkscape) Color is used for the text in Inkscape, but the package 'color.sty' is not loaded}%
    \renewcommand\color[2][]{}%
  }%
  \providecommand\transparent[1]{%
    \errmessage{(Inkscape) Transparency is used (non-zero) for the text in Inkscape, but the package 'transparent.sty' is not loaded}%
    \renewcommand\transparent[1]{}%
  }%
  \providecommand\rotatebox[2]{#2}%
  \newcommand*\fsize{\dimexpr\f@size pt\relax}%
  \newcommand*\lineheight[1]{\fontsize{\fsize}{#1\fsize}\selectfont}%
  \ifx\svgwidth\undefined%
    \setlength{\unitlength}{507.83352661bp}%
    \ifx\svgscale\undefined%
      \relax%
    \else%
      \setlength{\unitlength}{\unitlength * \real{\svgscale}}%
    \fi%
  \else%
    \setlength{\unitlength}{\svgwidth}%
  \fi%
  \global\let\svgwidth\undefined%
  \global\let\svgscale\undefined%
  \makeatother%
  \begin{picture}(1,0.28569601)%
    \lineheight{1}%
    \setlength\tabcolsep{0pt}%
    \put(0,0){\includegraphics[width=\unitlength]{A_qi.eps}}%
    \put(0.07486955,0.02679737){\makebox(0,0)[t]{\lineheight{1.25}\smash{\begin{tabular}[t]{c}−1\end{tabular}}}}%
    \put(0.15162973,0.02679737){\makebox(0,0)[t]{\lineheight{1.25}\smash{\begin{tabular}[t]{c}0\end{tabular}}}}%
    \put(0.22838989,0.02679737){\makebox(0,0)[t]{\lineheight{1.25}\smash{\begin{tabular}[t]{c}1\end{tabular}}}}%
    \put(0.15162973,0.00776586){\makebox(0,0)[t]{\lineheight{1.25}\smash{\begin{tabular}[t]{c}$\Re \{ S_{11} \} $\end{tabular}}}}%
    \put(0.05252521,0.05405102){\makebox(0,0)[rt]{\lineheight{1.25}\smash{\begin{tabular}[t]{r}−1.0\end{tabular}}}}%
    \put(0.05252521,0.09310533){\makebox(0,0)[rt]{\lineheight{1.25}\smash{\begin{tabular}[t]{r}−0.5\end{tabular}}}}%
    \put(0.05252521,0.13215963){\makebox(0,0)[rt]{\lineheight{1.25}\smash{\begin{tabular}[t]{r}0.0\end{tabular}}}}%
    \put(0.05252521,0.17121393){\makebox(0,0)[rt]{\lineheight{1.25}\smash{\begin{tabular}[t]{r}0.5\end{tabular}}}}%
    \put(0.05252521,0.21026823){\makebox(0,0)[rt]{\lineheight{1.25}\smash{\begin{tabular}[t]{r}1.0\end{tabular}}}}%
    \put(0.01541228,0.13686666){\rotatebox{90}{\makebox(0,0)[t]{\lineheight{1.25}\smash{\begin{tabular}[t]{c}$\Im \{ S_{11} \} $\end{tabular}}}}}%
    \put(0.31158098,0.02679737){\makebox(0,0)[t]{\lineheight{1.25}\smash{\begin{tabular}[t]{c}−1\end{tabular}}}}%
    \put(0.38834115,0.02679737){\makebox(0,0)[t]{\lineheight{1.25}\smash{\begin{tabular}[t]{c}0\end{tabular}}}}%
    \put(0.46510132,0.02679737){\makebox(0,0)[t]{\lineheight{1.25}\smash{\begin{tabular}[t]{c}1\end{tabular}}}}%
    \put(0.38834115,0.00776586){\makebox(0,0)[t]{\lineheight{1.25}\smash{\begin{tabular}[t]{c}$\Re \{ S_{11} \} $\end{tabular}}}}%
    \put(0.28923666,0.05405102){\makebox(0,0)[rt]{\lineheight{1.25}\smash{\begin{tabular}[t]{r}−1.0\end{tabular}}}}%
    \put(0.28923666,0.09310533){\makebox(0,0)[rt]{\lineheight{1.25}\smash{\begin{tabular}[t]{r}−0.5\end{tabular}}}}%
    \put(0.28923666,0.13215963){\makebox(0,0)[rt]{\lineheight{1.25}\smash{\begin{tabular}[t]{r}0.0\end{tabular}}}}%
    \put(0.28923666,0.17121393){\makebox(0,0)[rt]{\lineheight{1.25}\smash{\begin{tabular}[t]{r}0.5\end{tabular}}}}%
    \put(0.28923666,0.21026823){\makebox(0,0)[rt]{\lineheight{1.25}\smash{\begin{tabular}[t]{r}1.0\end{tabular}}}}%
    \put(0.59259826,0.02679737){\makebox(0,0)[t]{\lineheight{1.25}\smash{\begin{tabular}[t]{c}−1\end{tabular}}}}%
    \put(0.66935843,0.02679737){\makebox(0,0)[t]{\lineheight{1.25}\smash{\begin{tabular}[t]{c}0\end{tabular}}}}%
    \put(0.74611866,0.02679737){\makebox(0,0)[t]{\lineheight{1.25}\smash{\begin{tabular}[t]{c}1\end{tabular}}}}%
    \put(0.66935843,0.00776586){\makebox(0,0)[t]{\lineheight{1.25}\smash{\begin{tabular}[t]{c}$\Re \{ S_{11} \} $\end{tabular}}}}%
    \put(0.57025394,0.05405102){\makebox(0,0)[rt]{\lineheight{1.25}\smash{\begin{tabular}[t]{r}−1.0\end{tabular}}}}%
    \put(0.57025394,0.09310533){\makebox(0,0)[rt]{\lineheight{1.25}\smash{\begin{tabular}[t]{r}−0.5\end{tabular}}}}%
    \put(0.57025394,0.13215963){\makebox(0,0)[rt]{\lineheight{1.25}\smash{\begin{tabular}[t]{r}0.0\end{tabular}}}}%
    \put(0.57025394,0.17121393){\makebox(0,0)[rt]{\lineheight{1.25}\smash{\begin{tabular}[t]{r}0.5\end{tabular}}}}%
    \put(0.57025394,0.21026823){\makebox(0,0)[rt]{\lineheight{1.25}\smash{\begin{tabular}[t]{r}1.0\end{tabular}}}}%
    \put(0.53314103,0.13686662){\rotatebox{90}{\makebox(0,0)[t]{\lineheight{1.25}\smash{\begin{tabular}[t]{c}$\Im \{ S_{11} \} $\end{tabular}}}}}%
    \put(0.82930975,0.02679737){\makebox(0,0)[t]{\lineheight{1.25}\smash{\begin{tabular}[t]{c}−1\end{tabular}}}}%
    \put(0.90606991,0.02679737){\makebox(0,0)[t]{\lineheight{1.25}\smash{\begin{tabular}[t]{c}0\end{tabular}}}}%
    \put(0.98283008,0.02679737){\makebox(0,0)[t]{\lineheight{1.25}\smash{\begin{tabular}[t]{c}1\end{tabular}}}}%
    \put(0.90606991,0.00776586){\makebox(0,0)[t]{\lineheight{1.25}\smash{\begin{tabular}[t]{c}$\Re \{ S_{11} \} $\end{tabular}}}}%
    \put(0.80696536,0.05405102){\makebox(0,0)[rt]{\lineheight{1.25}\smash{\begin{tabular}[t]{r}−1.0\end{tabular}}}}%
    \put(0.80696536,0.09310533){\makebox(0,0)[rt]{\lineheight{1.25}\smash{\begin{tabular}[t]{r}−0.5\end{tabular}}}}%
    \put(0.80696536,0.13215963){\makebox(0,0)[rt]{\lineheight{1.25}\smash{\begin{tabular}[t]{r}0.0\end{tabular}}}}%
    \put(0.80696536,0.17121393){\makebox(0,0)[rt]{\lineheight{1.25}\smash{\begin{tabular}[t]{r}0.5\end{tabular}}}}%
    \put(0.80696536,0.21026823){\makebox(0,0)[rt]{\lineheight{1.25}\smash{\begin{tabular}[t]{r}1.0\end{tabular}}}}%
    \put(0.04723489,0.28042197){\color[rgb]{0,0,0}\makebox(0,0)[lt]{\begin{minipage}{0.45204683\unitlength}\centering Resonator A\end{minipage}}}%
    \put(0.56021308,0.28031814){\color[rgb]{0,0,0}\makebox(0,0)[lt]{\begin{minipage}{0.45204683\unitlength}\centering Resonator C\end{minipage}}}%
    \put(0.15285455,0.25664984){\color[rgb]{0,0,1}\makebox(0,0)[lt]{\begin{minipage}{0.10713247\unitlength}\centering K\end{minipage}}}%
    \put(0.17943807,0.25644215){\color[rgb]{0,0.50196078,0}\makebox(0,0)[lt]{\begin{minipage}{0.10713247\unitlength}\centering G\end{minipage}}}%
    \put(0.24036947,0.25616525){\color[rgb]{1,0,0}\makebox(0,0)[lt]{\begin{minipage}{0.10713247\unitlength}\centering G + ThO$_2$\end{minipage}}}%
    \put(0.66958374,0.25664984){\color[rgb]{0,0,1}\makebox(0,0)[lt]{\begin{minipage}{0.10713247\unitlength}\centering K\end{minipage}}}%
    \put(0.69616727,0.25644215){\color[rgb]{0,0.50196078,0}\makebox(0,0)[lt]{\begin{minipage}{0.10713247\unitlength}\centering G\end{minipage}}}%
    \put(0.75709866,0.25616525){\color[rgb]{1,0,0}\makebox(0,0)[lt]{\begin{minipage}{0.10713247\unitlength}\centering G + ThO$_2$\end{minipage}}}%
  \end{picture}%
\endgroup%

%% file: A_uncert.eps_tex
\begingroup%
  \makeatletter%
  \providecommand\color[2][]{%
    \errmessage{(Inkscape) Color is used for the text in Inkscape, but the package 'color.sty' is not loaded}%
    \renewcommand\color[2][]{}%
  }%
  \providecommand\transparent[1]{%
    \errmessage{(Inkscape) Transparency is used (non-zero) for the text in Inkscape, but the package 'transparent.sty' is not loaded}%
    \renewcommand\transparent[1]{}%
  }%
  \providecommand\rotatebox[2]{#2}%
  \newcommand*\fsize{\dimexpr\f@size pt\relax}%
  \newcommand*\lineheight[1]{\fontsize{\fsize}{#1\fsize}\selectfont}%
  \ifx\svgwidth\undefined%
    \setlength{\unitlength}{519.40002441bp}%
    \ifx\svgscale\undefined%
      \relax%
    \else%
      \setlength{\unitlength}{\unitlength * \real{\svgscale}}%
    \fi%
  \else%
    \setlength{\unitlength}{\svgwidth}%
  \fi%
  \global\let\svgwidth\undefined%
  \global\let\svgscale\undefined%
  \makeatother%
  \begin{picture}(1,0.28663825)%
    \lineheight{1}%
    \setlength\tabcolsep{0pt}%
    \put(0,0){\includegraphics[width=\unitlength]{A_uncert.eps}}%
    \put(0.09117519,0.04530481){\makebox(0,0)[t]{\lineheight{1.25}\smash{\begin{tabular}[t]{c}7.445\end{tabular}}}}%
    \put(0.18315603,0.04530481){\makebox(0,0)[t]{\lineheight{1.25}\smash{\begin{tabular}[t]{c}7.455\end{tabular}}}}%
    \put(0.27513685,0.04530481){\makebox(0,0)[t]{\lineheight{1.25}\smash{\begin{tabular}[t]{c}7.465\end{tabular}}}}%
    \put(0.36711771,0.04530481){\makebox(0,0)[t]{\lineheight{1.25}\smash{\begin{tabular}[t]{c}7.475\end{tabular}}}}%
    \put(0.45909856,0.04530481){\makebox(0,0)[t]{\lineheight{1.25}\smash{\begin{tabular}[t]{c}7.485\end{tabular}}}}%
    \put(0.27968746,0.01812531){\makebox(0,0)[t]{\lineheight{1.25}\smash{\begin{tabular}[t]{c}$f$(GHz)\end{tabular}}}}%
    \put(0.06674047,0.08519862){\makebox(0,0)[rt]{\lineheight{1.25}\smash{\begin{tabular}[t]{r}−2\end{tabular}}}}%
    \put(0.06674047,0.14417204){\makebox(0,0)[rt]{\lineheight{1.25}\smash{\begin{tabular}[t]{r}0\end{tabular}}}}%
    \put(0.06674047,0.20314549){\makebox(0,0)[rt]{\lineheight{1.25}\smash{\begin{tabular}[t]{r}2\end{tabular}}}}%
    \put(0.04055099,0.15786678){\rotatebox{90}{\makebox(0,0)[t]{\lineheight{1.25}\smash{\begin{tabular}[t]{c}arg\{$S_{11}$\} (rads)\end{tabular}}}}}%
    \put(0.37747638,0.20978467){\makebox(0,0)[lt]{\lineheight{1.25}\smash{\begin{tabular}[t]{l}Eq.~\eqref{arctanfit}\end{tabular}}}}%
    \put(0.37619898,0.22875882){\makebox(0,0)[lt]{\lineheight{1.25}\smash{\begin{tabular}[t]{l}raw data\end{tabular}}}}%
    \put(0.52044502,0.27120385){\color[rgb]{1,0.49803922,0.05490196}\makebox(0,0)[t]{\lineheight{1.25}\smash{\begin{tabular}[t]{c}Resonator B\end{tabular}}}}%
    \put(0.01630061,0.2605771){\color[rgb]{0,0,0}\makebox(0,0)[lt]{\begin{minipage}{0.11061393\unitlength}\raggedright \textbf{a}\end{minipage}}}%
    \put(0.45958198,0.2605772){\color[rgb]{0,0,0}\makebox(0,0)[lt]{\begin{minipage}{0.11061393\unitlength}\centering \textbf{b}\end{minipage}}}%
    \put(0.5843923,0.04676047){\makebox(0,0)[t]{\lineheight{1.25}\smash{\begin{tabular}[t]{c}−4\end{tabular}}}}%
    \put(0.67282143,0.04676047){\makebox(0,0)[t]{\lineheight{1.25}\smash{\begin{tabular}[t]{c}−2\end{tabular}}}}%
    \put(0.76125051,0.04676047){\makebox(0,0)[t]{\lineheight{1.25}\smash{\begin{tabular}[t]{c}0\end{tabular}}}}%
    \put(0.84967964,0.04676047){\makebox(0,0)[t]{\lineheight{1.25}\smash{\begin{tabular}[t]{c}2\end{tabular}}}}%
    \put(0.93810865,0.04676047){\makebox(0,0)[t]{\lineheight{1.25}\smash{\begin{tabular}[t]{c}4\end{tabular}}}}%
    \put(0.7605211,0.02996297){\makebox(0,0)[t]{\lineheight{1.25}\smash{\begin{tabular}[t]{c}$\varphi \prime$ (millirads)\end{tabular}}}}%
    \put(0.54961675,0.06017333){\makebox(0,0)[rt]{\lineheight{1.25}\smash{\begin{tabular}[t]{r}0\end{tabular}}}}%
    \put(0.54961675,0.09177983){\makebox(0,0)[rt]{\lineheight{1.25}\smash{\begin{tabular}[t]{r}10\end{tabular}}}}%
    \put(0.54961675,0.1233864){\makebox(0,0)[rt]{\lineheight{1.25}\smash{\begin{tabular}[t]{r}20\end{tabular}}}}%
    \put(0.54961675,0.15499296){\makebox(0,0)[rt]{\lineheight{1.25}\smash{\begin{tabular}[t]{r}30\end{tabular}}}}%
    \put(0.54961675,0.18659955){\makebox(0,0)[rt]{\lineheight{1.25}\smash{\begin{tabular}[t]{r}40\end{tabular}}}}%
    \put(0.54961675,0.21820614){\makebox(0,0)[rt]{\lineheight{1.25}\smash{\begin{tabular}[t]{r}50\end{tabular}}}}%
    \put(0.51590591,0.15716002){\rotatebox{90}{\makebox(0,0)[t]{\lineheight{1.25}\smash{\begin{tabular}[t]{c}Counts\end{tabular}}}}}%
    \put(0.59114748,0.24387959){\color[rgb]{0,0,0}\makebox(0,0)[lt]{\begin{minipage}{0.07796728\unitlength}\raggedright  $\pm \Delta \varphi$\end{minipage}}}%
  \end{picture}%
\endgroup%

%% file: A_histograms.eps_tex
\begingroup%
  \makeatletter%
  \providecommand\color[2][]{%
    \errmessage{(Inkscape) Color is used for the text in Inkscape, but the package 'color.sty' is not loaded}%
    \renewcommand\color[2][]{}%
  }%
  \providecommand\transparent[1]{%
    \errmessage{(Inkscape) Transparency is used (non-zero) for the text in Inkscape, but the package 'transparent.sty' is not loaded}%
    \renewcommand\transparent[1]{}%
  }%
  \providecommand\rotatebox[2]{#2}%
  \newcommand*\fsize{\dimexpr\f@size pt\relax}%
  \newcommand*\lineheight[1]{\fontsize{\fsize}{#1\fsize}\selectfont}%
  \ifx\svgwidth\undefined%
    \setlength{\unitlength}{531.01000977bp}%
    \ifx\svgscale\undefined%
      \relax%
    \else%
      \setlength{\unitlength}{\unitlength * \real{\svgscale}}%
    \fi%
  \else%
    \setlength{\unitlength}{\svgwidth}%
  \fi%
  \global\let\svgwidth\undefined%
  \global\let\svgscale\undefined%
  \makeatother%
  \begin{picture}(1,0.50210918)%
    \lineheight{1}%
    \setlength\tabcolsep{0pt}%
    \put(0,0){\includegraphics[width=\unitlength]{A_histograms.eps}}%
    \put(0.15503619,0.26998268){\makebox(0,0)[t]{\lineheight{1.25}\smash{\begin{tabular}[t]{c}100\end{tabular}}}}%
    \put(0.42181082,0.26998268){\makebox(0,0)[t]{\lineheight{1.25}\smash{\begin{tabular}[t]{c}1000\end{tabular}}}}%
    \put(0.06399476,0.2904782){\makebox(0,0)[rt]{\lineheight{1.25}\smash{\begin{tabular}[t]{r}10$^{-4}$\end{tabular}}}}%
    \put(0.06399476,0.34475186){\makebox(0,0)[rt]{\lineheight{1.25}\smash{\begin{tabular}[t]{r}10$^{-3}$\end{tabular}}}}%
    \put(0.06399476,0.39902553){\makebox(0,0)[rt]{\lineheight{1.25}\smash{\begin{tabular}[t]{r}10$^{-2}$\end{tabular}}}}%
    \put(0.06399476,0.45329919){\makebox(0,0)[rt]{\lineheight{1.25}\smash{\begin{tabular}[t]{r}10$^{-1}$\end{tabular}}}}%
    \put(0.01728719,0.37853901){\rotatebox{90}{\makebox(0,0)[t]{\lineheight{1.25}\smash{\begin{tabular}[t]{c}Counts / s\end{tabular}}}}}%
    \put(0.2884235,0.47865711){\makebox(0,0)[t]{\lineheight{1.25}\smash{\begin{tabular}[t]{c}\textbf{R}, glue: vacuum grease\end{tabular}}}}%
    \put(0.36479948,0.44325288){\makebox(0,0)[lt]{\lineheight{1.25}\smash{\begin{tabular}[t]{l}Resonator A\end{tabular}}}}%
    \put(0.36479948,0.42190442){\makebox(0,0)[lt]{\lineheight{1.25}\smash{\begin{tabular}[t]{l}Resonator B\end{tabular}}}}%
    \put(0.36479948,0.40055595){\makebox(0,0)[lt]{\lineheight{1.25}\smash{\begin{tabular}[t]{l}Resonator C\end{tabular}}}}%
    \put(0.63846934,0.26998268){\makebox(0,0)[t]{\lineheight{1.25}\smash{\begin{tabular}[t]{c}100\end{tabular}}}}%
    \put(0.90524398,0.26998268){\makebox(0,0)[t]{\lineheight{1.25}\smash{\begin{tabular}[t]{c}1000\end{tabular}}}}%
    \put(0.54742792,0.2904782){\makebox(0,0)[rt]{\lineheight{1.25}\smash{\begin{tabular}[t]{r}10$^{-4}$\end{tabular}}}}%
    \put(0.54742792,0.34475186){\makebox(0,0)[rt]{\lineheight{1.25}\smash{\begin{tabular}[t]{r}10$^{-3}$\end{tabular}}}}%
    \put(0.54742792,0.39902553){\makebox(0,0)[rt]{\lineheight{1.25}\smash{\begin{tabular}[t]{r}10$^{-2}$\end{tabular}}}}%
    \put(0.54742792,0.45329919){\makebox(0,0)[rt]{\lineheight{1.25}\smash{\begin{tabular}[t]{r}10$^{-1}$\end{tabular}}}}%
    \put(0.77185663,0.47865711){\makebox(0,0)[t]{\lineheight{1.25}\smash{\begin{tabular}[t]{c}\textbf{G}, Pb shield removed\end{tabular}}}}%
    \put(0.15503619,0.03510201){\makebox(0,0)[t]{\lineheight{1.25}\smash{\begin{tabular}[t]{c}100\end{tabular}}}}%
    \put(0.42181082,0.03510201){\makebox(0,0)[t]{\lineheight{1.25}\smash{\begin{tabular}[t]{c}1000\end{tabular}}}}%
    \put(0.2884235,0.01690118){\makebox(0,0)[t]{\lineheight{1.25}\smash{\begin{tabular}[t]{c}$10^6 \times \delta x_\mathrm{QP}$\end{tabular}}}}%
    \put(0.06399476,0.05559754){\makebox(0,0)[rt]{\lineheight{1.25}\smash{\begin{tabular}[t]{r}10$^{-4}$\end{tabular}}}}%
    \put(0.06399476,0.1098712){\makebox(0,0)[rt]{\lineheight{1.25}\smash{\begin{tabular}[t]{r}10$^{-3}$\end{tabular}}}}%
    \put(0.06399476,0.16414486){\makebox(0,0)[rt]{\lineheight{1.25}\smash{\begin{tabular}[t]{r}10$^{-2}$\end{tabular}}}}%
    \put(0.06399476,0.21841852){\makebox(0,0)[rt]{\lineheight{1.25}\smash{\begin{tabular}[t]{r}10$^{-1}$\end{tabular}}}}%
    \put(0.01728717,0.14365832){\rotatebox{90}{\makebox(0,0)[t]{\lineheight{1.25}\smash{\begin{tabular}[t]{c}Counts / s\end{tabular}}}}}%
    \put(0.2884235,0.24377642){\makebox(0,0)[t]{\lineheight{1.25}\smash{\begin{tabular}[t]{c}\textbf{R}, glue: Ag paste\end{tabular}}}}%
    \put(0.63846934,0.03510201){\makebox(0,0)[t]{\lineheight{1.25}\smash{\begin{tabular}[t]{c}100\end{tabular}}}}%
    \put(0.90524398,0.03510201){\makebox(0,0)[t]{\lineheight{1.25}\smash{\begin{tabular}[t]{c}1000\end{tabular}}}}%
    \put(0.77185663,0.01690118){\makebox(0,0)[t]{\lineheight{1.25}\smash{\begin{tabular}[t]{c}$10^6 \times \delta x_\mathrm{QP}$\end{tabular}}}}%
    \put(0.54742792,0.05559754){\makebox(0,0)[rt]{\lineheight{1.25}\smash{\begin{tabular}[t]{r}10$^{-4}$\end{tabular}}}}%
    \put(0.54742792,0.1098712){\makebox(0,0)[rt]{\lineheight{1.25}\smash{\begin{tabular}[t]{r}10$^{-3}$\end{tabular}}}}%
    \put(0.54742792,0.16414486){\makebox(0,0)[rt]{\lineheight{1.25}\smash{\begin{tabular}[t]{r}10$^{-2}$\end{tabular}}}}%
    \put(0.54742792,0.21841852){\makebox(0,0)[rt]{\lineheight{1.25}\smash{\begin{tabular}[t]{r}10$^{-1}$\end{tabular}}}}%
    \put(0.77185663,0.24377642){\makebox(0,0)[t]{\lineheight{1.25}\smash{\begin{tabular}[t]{c}\textbf{G}, Pb shield removed + Th0$_2$ source\end{tabular}}}}%
    \put(0.07176075,0.39356018){\color[rgb]{0,0,0}\makebox(0,0)[lt]{\begin{minipage}{0.63848247\unitlength}\centering Binning: $5 \times 10^{-5}$\end{minipage}}}%
  \end{picture}%
\endgroup%

%% file: A_simu.eps_tex
\begingroup%
  \makeatletter%
  \providecommand\color[2][]{%
    \errmessage{(Inkscape) Color is used for the text in Inkscape, but the package 'color.sty' is not loaded}%
    \renewcommand\color[2][]{}%
  }%
  \providecommand\transparent[1]{%
    \errmessage{(Inkscape) Transparency is used (non-zero) for the text in Inkscape, but the package 'transparent.sty' is not loaded}%
    \renewcommand\transparent[1]{}%
  }%
  \providecommand\rotatebox[2]{#2}%
  \newcommand*\fsize{\dimexpr\f@size pt\relax}%
  \newcommand*\lineheight[1]{\fontsize{\fsize}{#1\fsize}\selectfont}%
  \ifx\svgwidth\undefined%
    \setlength{\unitlength}{520.26361084bp}%
    \ifx\svgscale\undefined%
      \relax%
    \else%
      \setlength{\unitlength}{\unitlength * \real{\svgscale}}%
    \fi%
  \else%
    \setlength{\unitlength}{\svgwidth}%
  \fi%
  \global\let\svgwidth\undefined%
  \global\let\svgscale\undefined%
  \makeatother%
  \begin{picture}(1,0.26147941)%
    \lineheight{1}%
    \setlength\tabcolsep{0pt}%
    \put(0,0){\includegraphics[width=\unitlength]{A_simu.eps}}%
    \put(0.07504473,0.02791412){\makebox(0,0)[t]{\lineheight{1.25}\smash{\begin{tabular}[t]{c}1\end{tabular}}}}%
    \put(0.21408324,0.02791412){\makebox(0,0)[t]{\lineheight{1.25}\smash{\begin{tabular}[t]{c}10\end{tabular}}}}%
    \put(0.3531217,0.02791412){\makebox(0,0)[t]{\lineheight{1.25}\smash{\begin{tabular}[t]{c}100\end{tabular}}}}%
    \put(0.4921602,0.02791412){\makebox(0,0)[t]{\lineheight{1.25}\smash{\begin{tabular}[t]{c}1000\end{tabular}}}}%
    \put(0.28360246,0.00938233){\makebox(0,0)[t]{\lineheight{1.25}\smash{\begin{tabular}[t]{c}$\delta E$ (eV)\end{tabular}}}}%
    \put(0.06408874,0.09060349){\makebox(0,0)[rt]{\lineheight{1.25}\smash{\begin{tabular}[t]{r}10$^{-3}$\end{tabular}}}}%
    \put(0.06408874,0.17108479){\makebox(0,0)[rt]{\lineheight{1.25}\smash{\begin{tabular}[t]{r}10$^{-2}$\end{tabular}}}}%
    \put(0.01161114,0.15113535){\rotatebox{90}{\makebox(0,0)[t]{\lineheight{1.25}\smash{\begin{tabular}[t]{c}Counts / s\end{tabular}}}}}%
    \put(0.56935182,0.02791412){\makebox(0,0)[t]{\lineheight{1.25}\smash{\begin{tabular}[t]{c}1\end{tabular}}}}%
    \put(0.70839032,0.02791412){\makebox(0,0)[t]{\lineheight{1.25}\smash{\begin{tabular}[t]{c}10\end{tabular}}}}%
    \put(0.84742882,0.02791412){\makebox(0,0)[t]{\lineheight{1.25}\smash{\begin{tabular}[t]{c}100\end{tabular}}}}%
    \put(0.98646731,0.02791412){\makebox(0,0)[t]{\lineheight{1.25}\smash{\begin{tabular}[t]{c}1000\end{tabular}}}}%
    \put(0.7779096,0.00938233){\makebox(0,0)[t]{\lineheight{1.25}\smash{\begin{tabular}[t]{c}$\delta E$ (keV)\end{tabular}}}}%
    \put(0.55551272,0.0528529){\makebox(0,0)[rt]{\lineheight{1.25}\smash{\begin{tabular}[t]{r}1\end{tabular}}}}%
    \put(0.55551272,0.1259641){\makebox(0,0)[rt]{\lineheight{1.25}\smash{\begin{tabular}[t]{r}10\end{tabular}}}}%
    \put(0.55551272,0.19907529){\makebox(0,0)[rt]{\lineheight{1.25}\smash{\begin{tabular}[t]{r}100\end{tabular}}}}%
    \put(0.52802248,0.15114661){\rotatebox{90}{\makebox(0,0)[t]{\lineheight{1.25}\smash{\begin{tabular}[t]{c}Counts\end{tabular}}}}}%
    \put(0.08775042,0.15262611){\color[rgb]{0.12156863,0.46666667,0.70588235}\makebox(0,0)[lt]{\begin{minipage}{0.35792833\unitlength}\raggedright $\langle \delta E \rangle \approx $ 20 eV\end{minipage}}}%
    \put(0.08775042,0.13100148){\color[rgb]{1,0.49803922,0.05490196}\makebox(0,0)[lt]{\begin{minipage}{0.35792833\unitlength}\raggedright $\langle \delta E \rangle \approx $ 160 eV\end{minipage}}}%
    \put(0.08775042,0.10937682){\color[rgb]{0.17254902,0.62745098,0.17254902}\makebox(0,0)[lt]{\begin{minipage}{0.35792833\unitlength}\raggedright $\langle \delta E \rangle \approx $ 10 eV\end{minipage}}}%
    \put(0.3778775,0.25105687){\color[rgb]{0.12156863,0.46666667,0.70588235}\makebox(0,0)[lt]{\begin{minipage}{0.35792833\unitlength}\raggedright Resonator A\end{minipage}}}%
    \put(0.3778775,0.22943223){\color[rgb]{1,0.49803922,0.05490196}\makebox(0,0)[lt]{\begin{minipage}{0.35792833\unitlength}\raggedright Resonator B\end{minipage}}}%
    \put(0.3778775,0.2078076){\color[rgb]{0.17254902,0.62745098,0.17254902}\makebox(0,0)[lt]{\begin{minipage}{0.35792833\unitlength}\raggedright Resonator C\end{minipage}}}%
    \put(0.58115972,0.13100148){\color[rgb]{0,0,0}\makebox(0,0)[lt]{\begin{minipage}{0.35792833\unitlength}\raggedright $\langle \delta E \rangle \approx $ 150 keV\end{minipage}}}%
    \put(-0.1904013,0.25335542){\color[rgb]{0,0,0}\makebox(0,0)[lt]{\begin{minipage}{0.4565499\unitlength}\centering \textbf{a}\end{minipage}}}%
    \put(0.296514,0.25335542){\color[rgb]{0,0,0}\makebox(0,0)[lt]{\begin{minipage}{0.4565499\unitlength}\centering \textbf{b}\end{minipage}}}%
    \put(0.58115972,0.15262611){\color[rgb]{0,0,0}\makebox(0,0)[lt]{\begin{minipage}{0.35792833\unitlength}\raggedright Energy deposited in the substrate\end{minipage}}}%
    \put(0.08775042,0.08886214){\color[rgb]{0,0,0}\makebox(0,0)[lt]{\begin{minipage}{0.35792833\unitlength}\raggedright Binning: 5 eV\end{minipage}}}%
    \put(0.58115972,0.08886214){\color[rgb]{0,0,0}\makebox(0,0)[lt]{\begin{minipage}{0.35792832\unitlength}\raggedright Binning: 5 keV\end{minipage}}}%
  \end{picture}%
\endgroup%